\pdfoutput=1
\documentclass{article}

\usepackage{PRIMEarxiv}

\usepackage[utf8]{inputenc} 
\usepackage[T1]{fontenc}    
\usepackage{hyperref}       
\usepackage{url}            
\usepackage{booktabs}       
\usepackage{amsfonts}       
\usepackage{nicefrac}       
\usepackage{microtype}      
\usepackage{lipsum}
\usepackage{fancyhdr}       
\usepackage{graphicx}       
\usepackage[numbers]{natbib}  
\graphicspath{{media/}}     
\usepackage{amsmath}

\title{Integrative characterization of the topography of V4 neural codes using deep learning approaches
}

\author{
  Yingjue Bian\textsuperscript{1} \and
  Tianye Wang\textsuperscript{2,3,4,5} \and
  Shiming Tang\textsuperscript{2,3,4,5} \and
  Tai-Sing Lee\textsuperscript{1,6} \\[0.5em]
  \textsuperscript{1}Carnegie Mellon University, Pittsburgh, PA 15213, USA\\
  \textsuperscript{2}School of Life Sciences, Peking University, Beijing 100871, China\\
  \textsuperscript{3}Peking-Tsinghua Center for Life Sciences, Beijing 100871, China\\
  \textsuperscript{4}IDG/McGovern Institute for Brain Research, Peking University, Beijing 100871, China\\
  \textsuperscript{5}Key Laboratory of Machine Perception (Ministry of Education), Peking University, Beijing 100871, China\\
  \textsuperscript{6}Computer Science Department and Neuroscience Institute,\\
  \hspace{1em}Carnegie Mellon University, Pittsburgh, PA 15213, USA\\[0.5em]
}

\begin{document}
\maketitle

\begin{abstract}
Area V4 is a mid-level stage of the macaque ventral visual stream, known to encode intermediate visual features such as color, curvature, corners, texture, 3D solid, and local form. Classical neurophysiological studies have typically examined these dimensions in isolation—contrasting V4 selectivity for shape versus texture, 3D solid surfaces versus 2D flat patterns, or object form versus texture. Yet how these tunings relate to one another within individual neurons, and how they are jointly organized across the cortical surface, remains unknown. For instance, does a neuron selective for 2D contour-defined shape prefer 3D solid surfaces or 2D flat surfaces? How are preferences for such heterogeneous attributes arranged in a common topographic map?  To address these questions, we leverage V4 “digital twins”—deep neural network models fitted to large-scale, wide-field calcium imaging data comprising tens of thousands of natural images. These digital twins allow us to systematically probe not only the stimulus dimensions explored in earlier studies, but also new, multidimensional stimulus sets that reveal additional aspects of the V4 code.  In this study, we find that neural pixels preferring 2D contour-defined shapes also tend to prefer 3D surface shape defined by shading or texture gradients and to object form. In contrast, pixels preferring 2D texture tend to prefer flat surfaces defined by uniform texture or reflectance. We propose this division of labor suggests that V4 might have decomposed the encoding of geometrical shape and surface appearance of visual stimuli in distinct populations of neurons, organized as interleaved distinct clusters in the V4 topological map.
\end{abstract}


\section{Introduction}

Primate vision proceeds through a hierarchical pathway with increasing feature complexity. In the ventral stream, area V4 occupies a mid-level hub, receiving from V1/V2 and projecting to IT. Once considered mainly color-selective, V4 encodes shape, color, texture, and depth; its neurons are selective for orientation and spatial frequency as well as complex shape properties, with preferences spanning color, disparity, and higher-order patterns \cite{Tanigawa2010V4,Roe2012UnifiedV4,NANDY20131102,Li202V4}. Functional imaging shows segregated clusters (e.g., color vs.\ orientation), revealing a spatially structured topography of feature representation in V4 \cite{Tanigawa2010V4,Tang2020V4}. Within this diverse repertoire, shape processing emerges as a prominent dimension of V4 function.

Unlike V1, which encodes simple oriented bars or gratings, V4 integrates multiple inputs to represent more complex boundary features. Many neurons are tuned to curvature (convex/concave) at specific angular positions relative to a shape’s center; they also respond to non-Cartesian gratings, curved patterns, and contour features such as corners and junctions \cite{Arcizet2009}. Thus, V4 extracts local outline parts (e.g., curve segments) that serve as building blocks for downstream object representations \cite{PASUPATHY2019199}. Consistent with this role, lesions or inactivation of V4 impair form discrimination, including shapes defined by color or texture cues \cite{WALSH199351,MERIGAN_2000}. Yet object appearance is determined not only by boundaries but also by surface properties, to which V4 is likewise sensitive.

In addition to shape processing, area V4 analyzes surface properties such as texture and shading. Early work reported selectivity for complex object features, including surface textures \cite{Kobatake1994ComplexObject,KimBairPasupathy2019a}. Many V4 neurons respond robustly to natural textures and can discriminate among them, implying sensitivity to higher-order image statistics beyond contrast or spatial frequency \cite{Arcizet2008NaturalTextures}; lesions to V4 impair texture discrimination even with intact basic vision and attention \cite{MERIGAN_2000}. Direct comparisons reveal a continuum from shape-preferring to texture-preferring cells, with partly independent tuning for boundary shape and surface micro-patterns within single neurons \cite{KimBairPasupathy2019a}. Shape stimuli often elicit stronger and more selective responses than textures, yet both attributes are represented \cite{KimBairPasupathy2019a}. This surface coding further extends beyond 2D appearance to cues that signal three-dimensional structure.

Beyond 2D shapes and textures, area V4 also encodes 3D form cues. Although long assumed to represent shape mainly in 2D terms given its intermediate stage \cite{Arcizet2009,SrinathEmondsWang2021a}, recent work shows explicit 3D information: approximately half of V4 neurons prefer “solid” stimuli defined by shading, specular highlights, or binocular disparity, responding more to volumetric surfaces than shape-matched flat controls \cite{SrinathEmondsWang2021a}. These solid–flat preferences suggest that analysis of 3D object structure is already well underway by the level of V4, foreshadowing its bridging role between early visual cortex and IT \cite{SrinathEmondsWang2021a}.

A central open question is how V4’s diverse feature representations are topographically organized. Converging evidence points to a spatially patterned architecture, with domains for color, orientation, curvature, and 3D shape rather than a fully intermixed map \cite{Tanigawa2010V4,SrinathEmondsWang2021a}. Optical imaging reveals alternating color- and orientation-preferring regions \cite{Tanigawa2010V4}, and newer imaging/recordings report spatially clustered preferences for flat versus solid surface features \cite{SrinathEmondsWang2021a}. Yet it remains unclear whether these local preferences compose an integrated global map or only patchy mosaics; unlike V1’s canonical columns, higher-order maps in V4 remain debated. Progress is limited by multiplex tuning across shape, texture, color, and depth, the high dimensionality of natural stimuli, and substantial cross-laboratory heterogeneity in stimulus sets, recording scales, and analysis pipelines \cite{KimBairPasupathy2019a,Okazawa2015V4TextureStats,Hu2020CurvatureV4,Zhang2023SFV4}. Resolving this question will require a framework that explicitly links feature tuning to cortical layout across paradigms, rather than accumulating additional isolated findings.

A practical route to overcome both the mapping problem and cross-lab heterogeneity is emerging from large-scale data and modern modeling. Wide-field calcium imaging of macaque V4 under natural images provides a common mesoscale substrate, while task-optimized and data-driven deep networks can align heterogeneous recordings in a shared representational space and generate testable predictions (“digital twins”) that bridge laboratories and modalities \cite{Wang2024,Henry2022CrowdingV1V4,Smith2020}. These resources enable quantitative cross-dataset comparisons of shape, texture, and 3D cues within a unified framework for V4 and offer a principled basis for evaluating whether local feature domains compose an integrated global map.

In this study, we leverage the pretrained digital-twin model of macaque V4 from \citet{Wang2024} to ask whether neural coding axes defined under matched operational criteria—\emph{texture–shape}, \emph{solid–flat shape}, and \emph{texture–object}—reflect a common representational dimension or instead constitute separable, interleaved maps. Specifically, we examine how the shape-versus texture-preferring populations described by \citet{KimBairPasupathy2019a} are arranged topographically on the cortical sheet; whether units that prefer boundary shape over texture are tuned to 2D flat surface shape or to 3D surface shape as characterized by \citet{SrinathEmondsWang2021a}; and to what extent texture-preferring units encode 2D versus 3D surface cues. Using the model’s mesoscale cortical coordinate frame, we quantify spatial coupling among these preferences, derive cross-dimension alignment metrics, and generate testable predictions for targeted physiology. Together, these analyses provide a unified account of how boundary form, surface statistics, and perceived solid are laid out across V4.

\section{Materials and Methods}
\label{sec:Materials and methods}

\subsection{Digital Twin of V4: Deep Learning Model}
In this study, we will perform "neurophysiological" experiments in silico on a digital twin of V4 to obtain a more integrative and comprehensive view of V4 neural codes to 2D and 3D shapes and textures. Our macaque V4 ``digital twin'' neural encoding model was a deep learnign CNN model trained on widefield calcium imaging from three macaques viewing natural images sampled from an ImageNet subset \cite{Imagenet}. Signals were spatially binned into $90\times90\,\mu\mathrm{m}$ pixels, and for each pixel the mean response across five recording days served as the supervisory target. The model accepts $100\times100$ RGB images and returns a normalized widefield response map corresponding to predicted per-pixel mean activity.

Conceptually, training followed a three-stage pipeline: (i) a high-capacity teacher obtained by transferring a generic vision backbone to neural data; (ii) knowledge distillation to a compact student using dense pseudo-targets on a large proxy corpus; and (iii) fine-tuning the student on the neural dataset with cross-validated objectives and weak spatial smoothness priors, followed by cross-animal registration to a common cortical frame. We did not modify the original architecture or training protocol; all implementation details (model components, hyperparameters, preprocessing, and evaluation) strictly follow the published description and are referenced to the original article \cite{Wang2024}.

\subsection{Assessing Shape and Texture Preference and Selectivity of V4 Digital Twins}
\begin{figure*}[ht]
\begin{center}
\includegraphics[width=\linewidth]{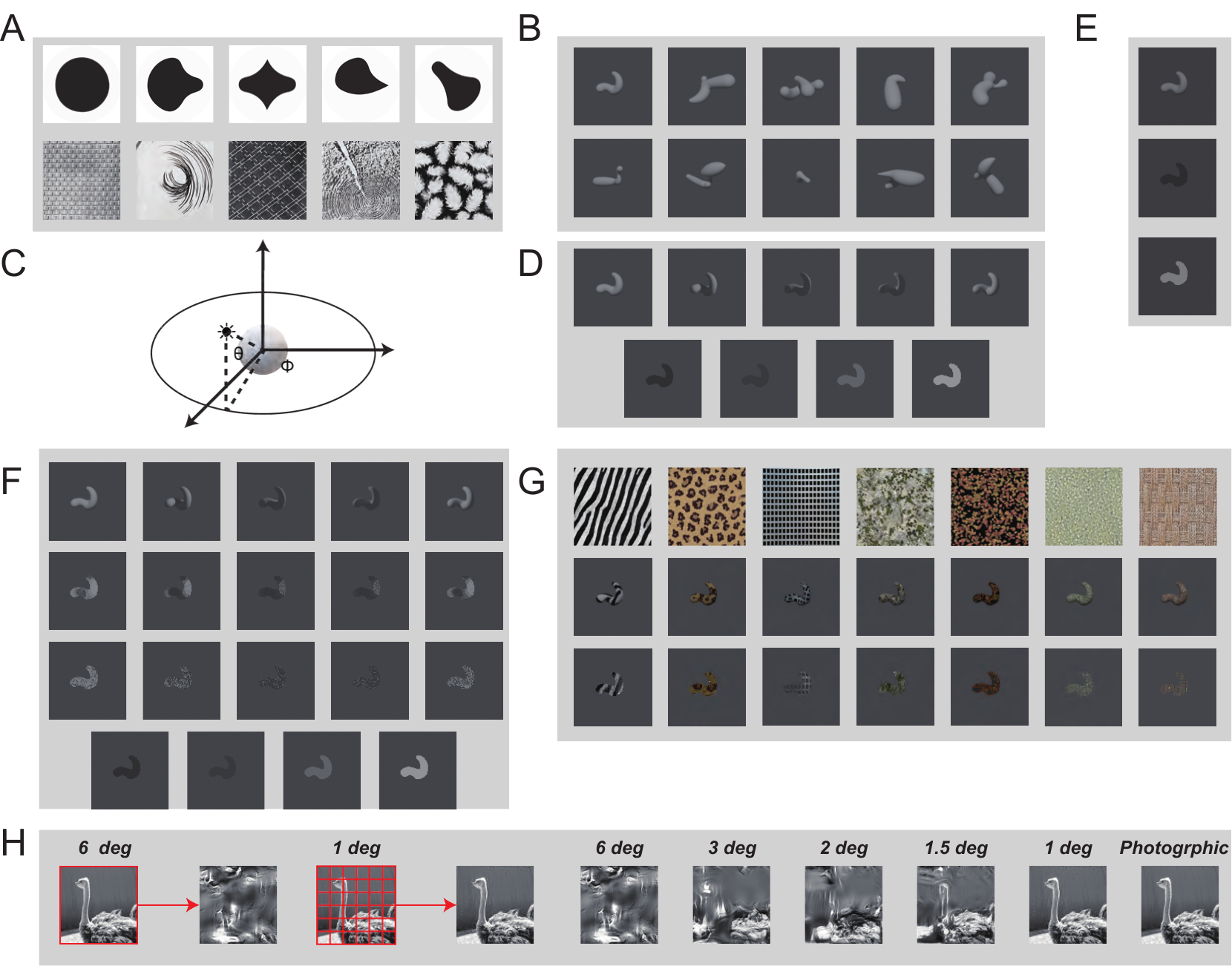}
\end{center}
\caption{Visual stimulus design. 
\textbf{(A)} Shape and texture stimuli reproduced from \cite{KimBairPasupathy2019a}. All stimuli were presented at four in-plane rotations in $45^{\circ}$ increments. 
\textbf{(B)} Solid/3D stimuli generated following \cite{SrinathEmondsWang2021a}. 
\textbf{(C)} Visualization of the lighting directions used for Blender renders. \textbf{(D)} Example stimulus family for a single object: solid renders under five lighting directions and flat controls at four contrast levels. 
\textbf{(E)} Additional stimulus examples following the design of \cite{SrinathEmondsWang2021a}. 
\textbf{(F)} Augmented variants for the same object: top row—solid; middle—“blob”; bottom—random controls. 
\textbf{(G)} Texture stimulus family: top row—web-sourced textures; middle—corresponding solid renders; bottom—flat controls.
\textbf{(H)} Texture-Object stimuli design similar to \cite{Lieber2024}, the objects image is a subset of object family from \cite{JohannesMehrer2021}.
} 
\label{fig:stimuli design}
\end{figure*}

\citet{KimBairPasupathy2019a} systematically characterized macaque V4 neurons’ selectivity to parametric 2D shapes and naturalistic textures with single-unit recordings and found that there are distinct populations of V4 neurons that are selective to 2D contour shapes or to texture independently. We first want to determine whether the units in the digital twin also exhibit these tuning properties. Second, since \citet{KimBairPasupathy2019a} recorded from one neuron at a time, it is clear how their shape neurons and texture neurons are organized topologically. Do the shape neurons and texture neurons cluster together into separate domains? The digital twin provides an opportunity for us to investigate how the shape neurons and texture neurons, as they define them, are organized across the cortical surface. 
 
We will test the digital twin with the same set of stimuli \citet{KimBairPasupathy2019a} tested in their study. Their shape family comprised 225 stimuli (30 stimuli orientated in 1, 2, 4 or 8 orientations in $45^{\circ}$) from \citet{PasupathyConnor2001a}, generated by systematically combining convex and concave boundary fragments. Since the original subset was not published \cite{KimBairPasupathy2019a}, we used the original full set from \citet{PasupathyConnor2001a}. The texture family comprised 112 images from \citet{Brodatz1966a}'s catalog. Each image was rotated by $0^\circ$, $90^\circ$, $180^\circ$, and $270^\circ$, yielding an augmented set of 816 stimuli. The example of these stimuli is shown in Figure~\ref{fig:stimuli design}A.

For size control, each stimulus was isotropically scaled so that its longer side equaled $3\bar{\sigma}$ of the target unit’s receptive field (RF), where the RF was estimated by fitting an elliptical Gaussian and $\bar{\sigma}=(\sigma_x+\sigma_y)/2$. For each neuronal column, the resized image was centered at the column’s estimated RF location on a uniform canvas and then presented to the model to obtain the predicted activation.

We quantified relative sensitivity to shape versus texture using a dispersion-based index computed from the within-category response variability: letting $\operatorname{STD}_{\text{shape}}$ and $\operatorname{STD}_{\text{texture}}$ denote the standard deviation (STD) of responses across the augmented shape and texture sets, respectively, the shape–texture sensitivity was
\begin{equation}
\label{eq: shape_texture_inde}
\mathrm{ST\text{-}sens} \;=\; \frac{\operatorname{STD}_{\text{shape}} - \operatorname{STD}_{\text{texture}}}{\operatorname{STD}_{\text{shape}} + \operatorname{STD}_{\text{texture}}}\,,
\end{equation}
optionally stabilized by adding a small $\varepsilon$ to the denominator. Positive values indicate greater dispersion (and thus sensitivity) for shapes, negative values for textures, and values near zero indicate comparable sensitivity to both. We calculate the texture preference index (TPI) with a similar pattern:
\begin{equation}
\label{eq:shape_texture_pref}
\operatorname{TPI} = \frac{\langle R_T\rangle - \langle R_S\rangle}{\langle R_T\rangle + \langle R_S\rangle}.
\end{equation}
Here, $\langle R_S\rangle$ and $\langle R_T\rangle$ denote the mean predicted responses to shape and texture stimuli, respectively. Concretely, we compute
$\langle R_S\rangle = \frac{1}{N}\sum_{i=1}^{N} R_{S,i}$ and
$\langle R_T\rangle = \frac{1}{M}\sum_{j=1}^{M} R_{T,j}$,
where $R_S$ and $R_T$ denote the predicted responses to individual shape and texture stimuli, and only responses larger than 0 (above baseline) are included in these averages considered as the valid response. Under this definition, positive TPI values indicate stronger mean responses to textures than shapes, negative values indicate stronger responses to shapes, and values near zero indicate no clear preference.

\subsection{Assessing Selectivity to 3D Solid Shapes}
\citet{SrinathEmondsWang2021a} investigated whether V4 neurons are selective to 3D solid surface structures. In this ground-breaking discovery, they showed that a significant number of neurons in V4 are sensitive and selective to 3D solids rendered in shading cues, as well as other cues. Because single-unit recording only allowed for testing a finite number of stimuli per neuron, they had to rely on an innovative genetic algorithm approach to search for the 3D solid shapes preferred by the neuron in a single recording session. The digital twin allows us to test each unit in the digital twin with an unlimited number of stimuli to gain a more definitive and comprehensive understanding of V4 tunings to 3D solid.
We generated a large family of solid 3D objects using their stimulus generation procedure as described in \cite{HungCarlsonConnor2012a}. Briefly, we first sampled a skeletal graph by randomly specifying limb configuration, limb lengths, centerline curvatures, and widths. The width profile of each limb was obtained by fitting a quadratic function to three independently sampled widths (two endpoints and the midpoint), ensuring smooth tapering. Limb junctions were then smoothed by applying a Gaussian kernel to the vertex positions and surface normals, yielding continuous curvature across branches.

In contrast to \citet{SrinathEmondsWang2021a}'s work, who used an online genetic algorithm to morph stimuli during electrophysiological recording, we pre-computed a large-scale library of solids (\(\sim\!3\times 10^{4}\) images, Figure~\ref{fig:stimuli design}B) that densely covers the parameter space. We will test digital twins of neurons with the entire library of solids and their flat counterparts. For each rendered object, the 2D correspondence information for downstream analyses was extracted from the rasterized object mask (i.e., silhouette/part-instance maps), providing pixel-wise alignment between the 3D model and its 2D projection.

It is important to recognize that neuronal responses in these studies are influenced by both the contour structure of objects and the 3D surface geometry. Exhaustively testing a full library of solid shapes helps reveal the preferred 3D structures. To dissociate selectivity for flat versus volumetric surfaces, \citet{SrinathEmondsWang2021a} factored out the contribution of contour shapes. They used stimuli in which the global contour outline associated with the preferred objects was held constant while comparing responses to solid 3D surfaces —rendered with various texture and shading cues—against flat versions of the surface defined by the same outer contour. In contrast, \citet{KimBairPasupathy2019a} compared neuronal responses to 2D contour shapes with responses to flat textured surfaces. Thus, whether the stimulus dimensions being compared and contrasted in these two studies are fully aligned remains an open question. Our study aims to leverage digital twins to compare cortical columns' preference profiles along these stimulus dimensions.

After defining the 3D parametric space, we rendered stimuli in Blender for presentation to the digital twin. The experiment comprised two matched paradigms: a shading-based set that manipulates illumination while holding geometry and viewing fixed, and a texture-based set that manipulates surface statistics under identical geometry and viewing.

\noindent\textbf{3D Solid from Shading Stimuli} \\For each object, camera, object pose, material, and background were fixed; only illumination varied. Point lights were placed on a cone with elevation $\theta=45^\circ$ and azimuth $\phi = k\times 60^\circ$ ($k\in\{0,\dots,5\}$) (Fig.~\ref{fig:stimuli design}B and C). One back-light configuration was excluded to avoid degenerate cues. Thus, image differences arose purely from shading (Fig.~\ref{fig:stimuli design}C).

\noindent\textbf{2D Flat controls.} Using the identical object mask, we aligned 2D patterns to the same nominal light direction but removed all 3D shading. We tested stimuli with uniform luminance within the contour shapes that match the mean luminance of the 3D solid based on shading stimuli, five level of luminances are tested (Fig.~\ref{fig:stimuli design}B). Inspired by \citet{Arcizet2009}, we introduced two additional two types of 2D flat surface stimuli, both with mean luminance within the outer contour mask matched to the uniform luminance 2D flat surfaces or the mean of each shading stimulus:  (i) stimuli with Gaussian noise (ii) stimuli with random blobs (Fig.~\ref{fig:stimuli design}F). These stimuli explicitly remove 3D object cues such as shading gradients, volumetric highlights and cast shadows, while preserving shape silhouette and the mean luminance of the stimuli, enabling direct comparison of 3D versus 2D drive.

\noindent\textbf{3D Solid from Texture and Shading Stimuli}\\
While shape-from-shading provides a strong cue for perceiving three-dimensional (3D) solids, we sought to test how sensitive this selectivity is to the specific cues that define 3D structure. \citet{SrinathEmondsWang2021a} examined a limited set of neurons using both shading- and stereo-disparity–defined stimuli and found their responses to be correlated. The digital twins enable us to extend this analysis by testing how the same neurons respond to 3D solids defined by texture cues, a stimulus condition not explored in \citet{SrinathEmondsWang2021a}. To generate the 3D shape-from-texture stimuli, we used the same object geometries, viewing poses, camera parameters, baseline material, and background as in the original experiments. Each object was rendered under a single frontal illumination at $45^\circ$ elevation (viewer-facing, as above) while applying photographic textures sampled from public libraries as PBR materials (diffuse/roughness/normal maps when available), producing realistic surface appearance under matched lighting geometry (Fig.~\ref{fig:stimuli design}G). This condition does not isolate shading; instead, it augments the smooth shading gradients with high-frequency surface texture, bringing the stimuli closer to natural 3D surfaces and to our 2D texture stimuli. Motivated by the fact that prior V4 solid studies did not combine shading and texture cues in this way, we designed this “shading+texture” condition to test whether enriching shading with texture strengthens 3D signals and the correspondence between 2D texture and 3D solid representations in the V4 digital twin.

\noindent\textbf{2D Flat Textured Control.} From the corresponding 3D renders, we extracted the object mask and inset the source texture within that mask on the same background (Fig.~\ref{fig:stimuli design}G). Global luminance/contrast were lightly normalized to the associated 3D image to ensure photometric comparability. This design holds identity and viewing constant, varies only the presence of shape-dependent shading, and uses ecologically realistic textures to test solidness versus flatness under naturalistic surface statistics.



\noindent\textbf{Metrics for Solid and Flat selectivity.} 
To compare with \citet{SrinathEmondsWang2021a}, we followed their definition of the Solid--Flat Index (SFI) to quantify preference for shading-defined solid feature versus flat controls at each recording pixel $i$:

\begin{equation}
\label{eq:sfi_ed}
\mathrm{SFI}_{\mathrm{B}}^{i}
=
\frac{R_{\mathrm{3D}}^{i} - \max\!\big(R_{\mathrm{2D}}^{i}\big)}
     {\max\!\Big(R_{\mathrm{3D}}^{i},\, \max\!\big(R_{\mathrm{2D}}^{i}\big)\Big)},
\end{equation}

\noindent where {\em B}  denotes the stimuli family similar to \citet{SrinathEmondsWang2021a}, with shading stimuli generated from one illumination direction only,  and $R_{\mathrm{3D}}^{i}$ is the best response of site $i$ to shaded (solid) stimuli and $\max(R_{\mathrm{2D}}^{i})$ is the best response to any flat/contrast control. By construction, $\mathrm{SFI}\in[-1,1]$; positive values indicate solid-preferring sites and negative values indicate flat-preferring sites.

\noindent\textbf{Illumination Invariant Solid Preference.}
Since we can test more stimuli using the digital twin, we generalize their study by testing 3D solids with shading rendered based on multiple illumination directions to evaluate whether the selectivity to 3D solids depends on illumination direction. We used shading stimuli based on five distinct lighting directions (while holding geometry, pose, material, camera, and background fixed) and expanded the flat controls to four matched uniform-contrast levels. For each site $i$, we then computed an SFI using the maxima across the full set of shaded versus contrast controls:
\begin{equation}
\label{eq:sfi_dense}
\mathrm{SFI}_{\mathrm{M}}^{i}
=
\frac{\max R_{\mathrm{3D}}^{i}\;-\;\max R_{\mathrm{2D}}^{i}}
     {\max\!\big(\max R_{\mathrm{3D}}^{i},\ \max R_{\mathrm{2D}}^{i}\big)},
\end{equation}
where $M$ denotes denser illumination stimuli, and $\max R_{\mathrm{3D}}^{i}$ and $\max R_{\mathrm{2D}}^{i}$ denote the best responses of site $i$ over the enumerated shaded and uniform-contrast conditions, respectively. The population distribution of $\mathrm{SFI}_{\mathrm{M}}$ remained approximately centered with a slight right tail, consistent with a mixture of solid- and flat-preferring sites under denser lighting. Projecting $\mathrm{SFI}_{\mathrm{M}}$ to the cortical surface revealed interleaved, spatially clustered \emph{solid} and \emph{flat} domains. 

\noindent\textbf{Solid Preference based on Shading and Noise Textures.}
The 2D control of the shading stimuli is basically uniform luminance within the surface, devoid of any features, which might lead to a confounded weak response. We added uniform noises and blobs, as developed in \cite{Arcizet2009}, the 2D controls to mitigate this effect. 

For each lighting direction, we pooled \emph{all} stimuli comprising (i) shaded 3D renders, (ii) four levels of uniform-contrast 2D controls, and (iii) two additional 2D controls: \textit{Blob} (a localized luminance patch placed within the object mask on the nominal highlight side) and \textit{Random-Noise} (texture-like variation confined to the mask). Within each illumination, solid preference was quantified as
\begin{equation}
\label{eq:sfi_all}
\mathrm{SFI}_{\mathrm{S}}^{i}
=
\frac{R_{\mathrm{3D}}^{i}-\max\!\big(R_{\mathrm{2D}}^{i}\big)}
     {\max\!\Big(R_{\mathrm{3D}}^{i},\,\max\!\big(R_{\mathrm{2D}}^{i}\big)\Big)},
\end{equation}
where $S$ indicate shading direction, and $R_{\mathrm{3D}}^{i}$ is the maximum response of site $i$ among shaded stimuli for that illumination, and $\max(R_{\mathrm{2D}}^{i})$ is the maximum among the corresponding 2D controls (Contrast, Blob, Noise). Cortical maps of $\mathrm{SFI}_{\mathrm{S}}^{i}$ exhibited interleaved solid/flat domains that were qualitatively consistent across all five lighting directions, indicating illumination-invariant mesoscale topology.

\medskip
\noindent\textbf{Solid Preference with Textured and Shaded Surfaces}
We further expanded the class of texture characteristics of the 3D solid surfaces by wrapping 5 distinct classes of textures (Figure~\ref{fig:stimuli design}G first row) onto the 3D solid stimuli to enhance the perception of 3D shapes (Figure~\ref{fig:stimuli design}G second row). 2D control are the uniform textures shown in Figure~\ref{fig:stimuli design}G third row. 
For every site $i$ and each texture pair $t$, we computed a within-pair SFI,
\begin{equation}
\label{eq:sfi_texture}
\mathrm{SFI}_{\mathrm{T}}^{i}
=
\frac{R_{\mathrm{3D}}^{i}-R_{\mathrm{2D}}^{i}}
     {\max\!\big(R_{\mathrm{3D}}^{i},\,R_{\mathrm{2D}}^{i}\big)}\,,
\end{equation}
where $T$ indicates texture family. Mapping $\mathrm{SFI}_{\mathrm{T}}^{i}$ to cortex revealed clustered, interleaved solid/flat domains for \emph{every} texture, and the domain layouts closely matched those obtained with the shading set, demonstrating robustness to changes in surface statistics.

\subsection{Assessing Texture and Object Sensitivity}
\citet{Lieber2024} extended this line of work by using naturalistic photographs and their Portilla–Simoncelli scrambled texture metamers to test, at the single-neuron level, whether V4 cells distinguish object-containing scenes from images that preserve only local texture statistics within spatially localized pooling regions. In their experiments, each neuron was probed with a small core set of 20 natural images (and their shifted, texture-matched counterparts), providing a powerful but necessarily limited view of photo–scramble discriminability. Inspired by this approach, we asked how this form of natural-image discriminability is organized topographically across V4. Whereas \citet{Lieber2024} primarily quantified discriminability using population decoders, here we focus on single-unit preference and its large-scale topographic layout: we quantify, for each model unit, a continuous photo–scramble modulation index and project this preference onto the cortical sheet, without explicitly training population-level classifiers. Using the digital twin, we can systematically stimulate every model unit with large numbers of photographs and their locally texture-matched versions, yielding a more general and comprehensive map of V4’s ability to respond preferentially to images that match the complex local statistics of natural photographs.

We therefore implemented an analogous texture to object contrast in the V4 digital twin to test how texture versus object tuning is organized across the cortical surface and how it relates to the other feature maps.
Building on this, we generated “scrambled” images using a refined Portilla–Simoncelli (P–S) texture synthesis framework that matches local pooling–region statistics \cite{Freeman2011, Portilla2000}. The stimuli of object photography is a subset of \citet{JohannesMehrer2021}.

To construct the scrambled stimuli, we arranged pooling regions as a square grid of smoothly overlapping fields that tile the image. Within each region, we measured a set of texture statistics. Concretely, we convolved the image with 32 oriented filters computed under periodic boundary conditions (4 orientations × 4 scales × 2 phases). For each filter, we extracted both the linear response and its amplitude to form “simple-cell–like” and “complex-cell–like” channels. We then computed pairwise products of these simple/complex responses across positions, orientations, and scales to obtain mid-level statistics, and took weighted averages of these products within each pooling region to yield a set of covariance statistics.

Synthesis proceeded by initializing with Gaussian white noise and iteratively adjusting pixels so that, within every subregion, the measured statistics matched their targets. For each spatial “shifted” subimage, we synthesized variants using pooling-window grids of 1×1, 2×2, 3×3, 4×4, and 6×6; when displayed, these corresponded to pooling-region visual angles of $6^{\circ}, 3^{\circ}, 2^{\circ}, 1.5^{\circ}, \text{and}, 1^{\circ}$, respectively. We also included the original (unscrambled) image, yielding six conditions in total (Fig.~\ref{fig:stimuli design}H).

To quantify the sensitivity of each neuron, we computed the modulation index (MI) using the following equation:
\begin{equation}
\label{eq:mi_tex_obj}
\operatorname{MI} = \frac{R_{\text{object}} -  R_T}{R_{\text{object}} +  R_T}.
\end{equation}
where $R_{\text{object}}$ is the mean response to object images;
$ R_T$ is the average response across the texture images.

\subsection{Correlation Analysis}
To assess the relationships among the three indices—solid vs.\ flat preference under shading, solid vs.\ flat preference under shading+texture, and the composite 2D shape--texture index—we computed all pairwise Pearson product--moment and Spearman rank correlations for each lighting direction, and reported effect sizes with corresponding interval estimates.

\section{Results}

\begin{figure*}[!t]
\begin{center}
\includegraphics[width=0.75\linewidth]{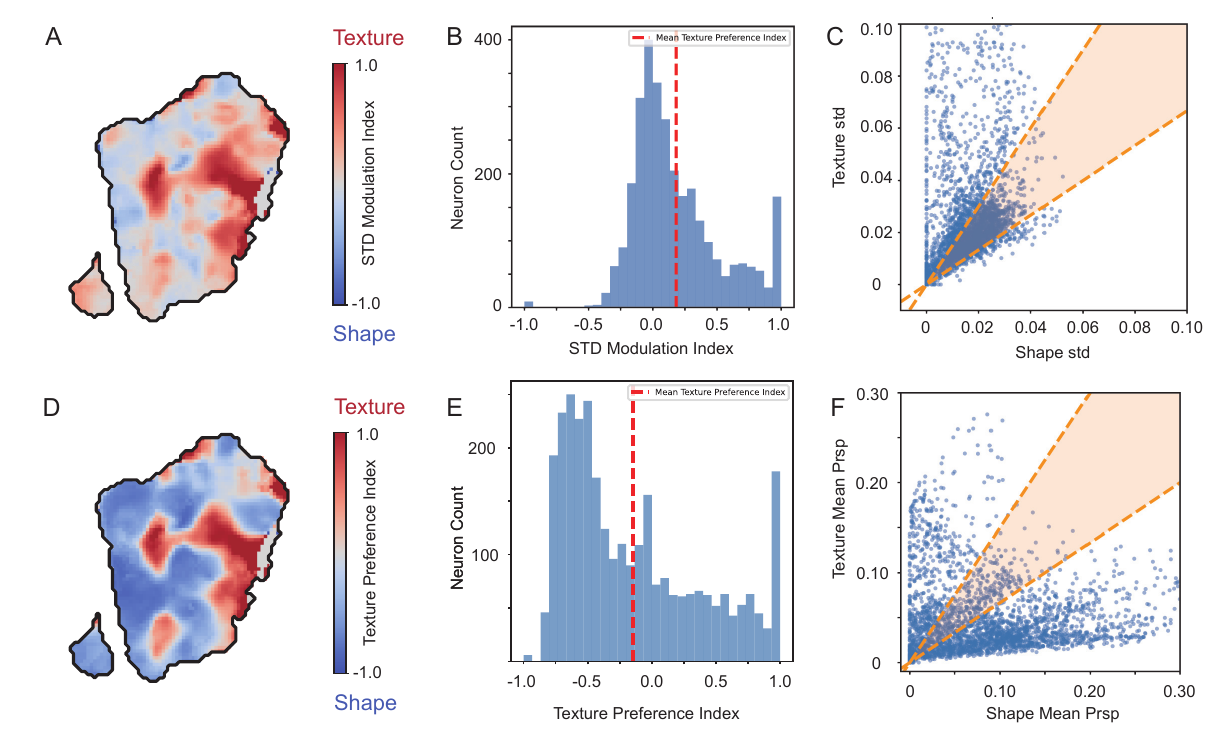}
\end{center}
\caption{\textbf{Texture--shape sensitivity within an example ROI.}
  \textbf{(A)} ROI map of the STD modulation index contrasting response variability under texture vs.\ shape manipulations; warmer colors indicate texture-dominated variance and cooler colors indicate shape-dominated variance. The black contour denotes the ROI boundary.
  \textbf{(B)} Distribution of the STD modulation index across units; the red dashed line marks the mean $=0.1824 $.
  \textbf{(C)} Per-unit scatter of standard deviations (x-axis: shape; y-axis: texture). The orange wedge denotes the band where the ratio satisfies a range of $[\frac{2}{3},\frac{3}{2}]$ ; points above/below the band show texture-/shape-dominated variability, respectively.
  \textbf{(D)} ROI map of the Texture Preference Index (TPI); warmer colors indicate stronger texture preference and cooler colors indicate stronger shape preference.
  \textbf{(E)} Distribution of the TPI across units; the red dashed line indicates the mean.
  Mean value is $-0.1468$
  \textbf{(F)} Per-unit scatter of mean response (x-axis: shape; y-axis: texture). The orange wedge denotes the band where the ratio satisfies a range of $[\frac{2}{3},\frac{3}{2}]$; points above/below the band show texture-/shape-dominated variability, respectively.
  } 
  \label{fig:texture_shape_roi}
\end{figure*}

\subsection{Preference and Selectivity to Shapes or Texture}
We quantified how responses varied under texture versus shape manipulations using a STD modulation index (calculated as Eq.~\ref{eq: shape_texture_inde}) and a TPI (calculated as Eq.~\ref{eq:shape_texture_pref}). Within the ROI, the STD modulation map exhibited locally coherent patches of texture-dominated and shape-dominated variability (Fig.~\ref{fig:texture_shape_roi}A), and the population histogram was broadly centered slightly above zero (mean $=0.1824$) with a modest skew, indicating a mild overall bias toward texture sensitivity  (Fig.~\ref{fig:texture_shape_roi}B). A per-unit scatter of texture versus shape STDs (Fig.~\ref{fig:texture_shape_roi}C) further isolated selective subpopulations: we defined a neutrality wedge by the ratio criterion \( \tfrac{2}{3} \le y/x \le \tfrac{3}{2} \) (texture STD \(y\), shape STD \(x\)), similar to \citet{KimBairPasupathy2019a}. Points outside this wedge showed clear sensitivity to texture (above) or shape (below), whereas points within the wedge exhibited comparable sensitivity to both manipulations. 

A complementary TPI revealed spatially clustered domains of texture- versus shape-biased units that closely paralleled the STD-based modulation pattern (Fig.~\ref{fig:texture_shape_roi}D). At the population level, the TP distribution was clearly non-Gaussian and broadly spread, with a mean slightly below zero (mean $=-0.1468$), indicating a mild overall bias toward shape responses while still encompassing a wide range of preferences (Fig.~\ref{fig:texture_shape_roi}E). Notably, a subset of neurons reached $\mathrm{TP}=1$, reflecting units that responded almost exclusively to texture stimuli and thus exhibited a strong texture preference. Consistent with this, plotting mean texture responses against mean shape responses showed a tight positive relationship but many units deviating substantially from the unity line, including a cluster with disproportionately larger texture responses (Fig.~\ref{fig:texture_shape_roi}F). Similarly, we marked a corresponding non-preference wedge by the constraint \( \tfrac{2}{3} \le \bar{R}_T / \bar{R}_S \le \tfrac{3}{2} \).

Relative to prior work reporting a larger fraction of units with $\mathrm{STD}_{\text{shape}}$ larger than $\mathrm{STD}_{\text{texture}}$ \cite{KimBairPasupathy2019a}, our dataset exhibits the converse bias, with more units showing \(\mathrm{STD}_{\text{texture}}\) smaller than \(\mathrm{STD}_{\text{shape}}\). Importantly, we did not summarize effects using the raw ratio \(\mathrm{STD}_{\text{shape}}/\mathrm{STD}_{\text{texture}}\), because in a nontrivial subset of units one of the response STDs is small (near zero), which yields heavy-tailed and potentially inflated ratios that are sensitive to measurement noise. Instead, all main analyses rely on the symmetric, bounded modulation index in Eq.~\ref{eq: shape_texture_inde}, which is scale-invariant and substantially more robust to small denominators. Under this robust metric, the texture-dominated variability remains evident at the population level within the ROI, despite condition-to-condition differences in stimulus ensembles and measurement noise. Consistent with this, the spatial patterns of STD modulation and texture preference (Fig.~\ref{fig:texture_shape_roi}A vs.~D) were strongly correlated (Pearson $r = 0.8316$, Spearman $\rho = 0.7232$), indicating that variance-based and mean-based indices capture largely overlapping texture- and shape-biased domains in the V4 digital twin.

\subsection{ Preferences for 3D Solid Shape or 2D Flat Surfaces }
\begin{figure*}[!t]
\begin{center}
\includegraphics[width=\linewidth]{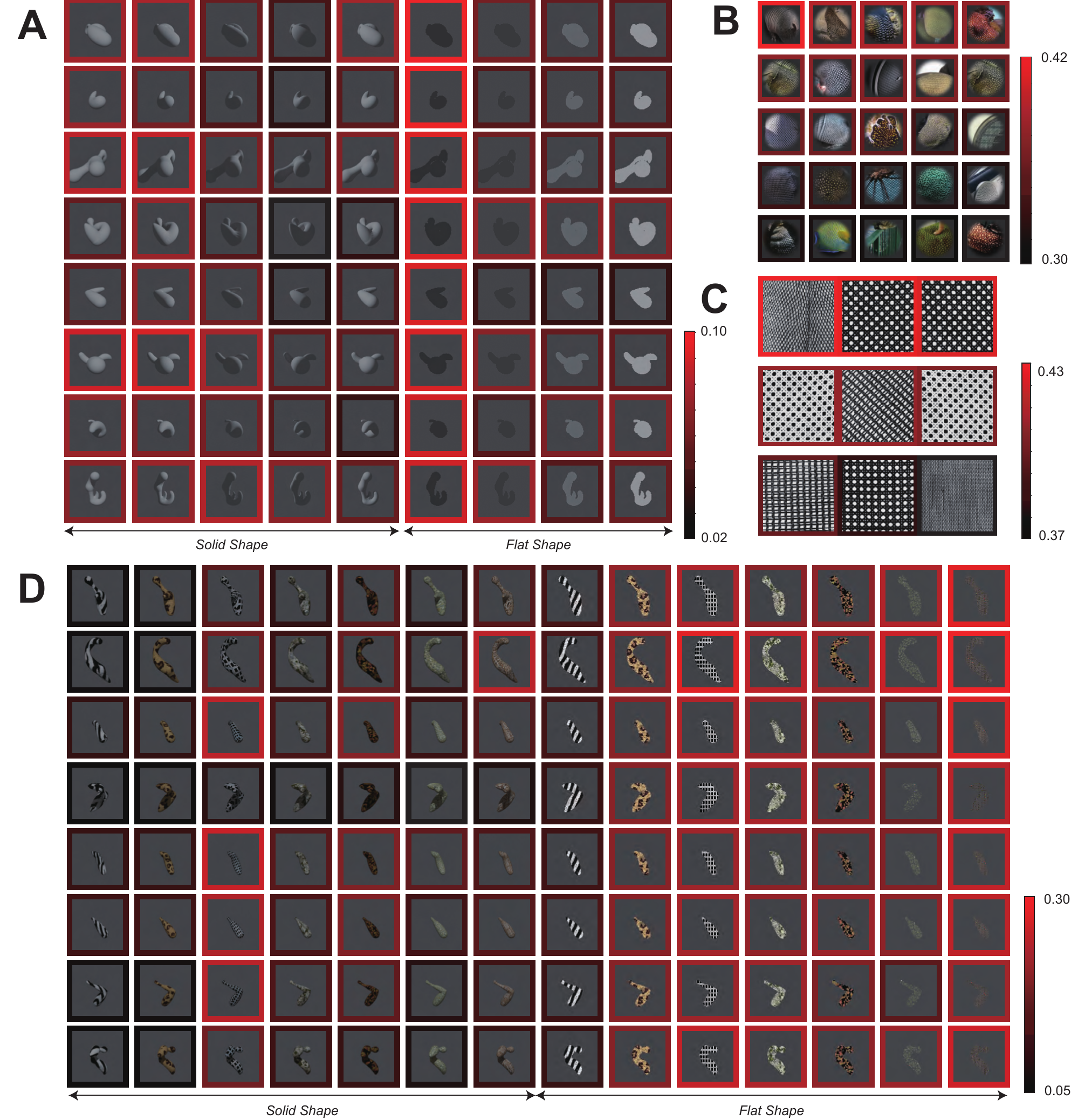}
\end{center}
\caption{\textbf{Area V4 Encodes flat-Shape Information-Illustrative Examples.}
    \textbf{(A)} Shading stimulus. Normalized response range: 0.02-0.10. Mean Solid-Flat Index (SFI) across lighting directions: -0.7752.
    \textbf{(B)} Preferred natural images. Response range: 0.30-0.42.
    \textbf{(C)} Preferred stimuli from the 2D shape and texture set of \citet{KimBairPasupathy2019a}. Response range: 0.37-0.43.
    \textbf{(D)} Texture render. Normalized response range: $0.02$-$0.10$. Mean SFI across seven textures: $-0.3930$}
  \label{fig:Neuron_1_example}
\end{figure*}

\begin{figure*}[!t]
\begin{center}
\includegraphics[width=\linewidth]{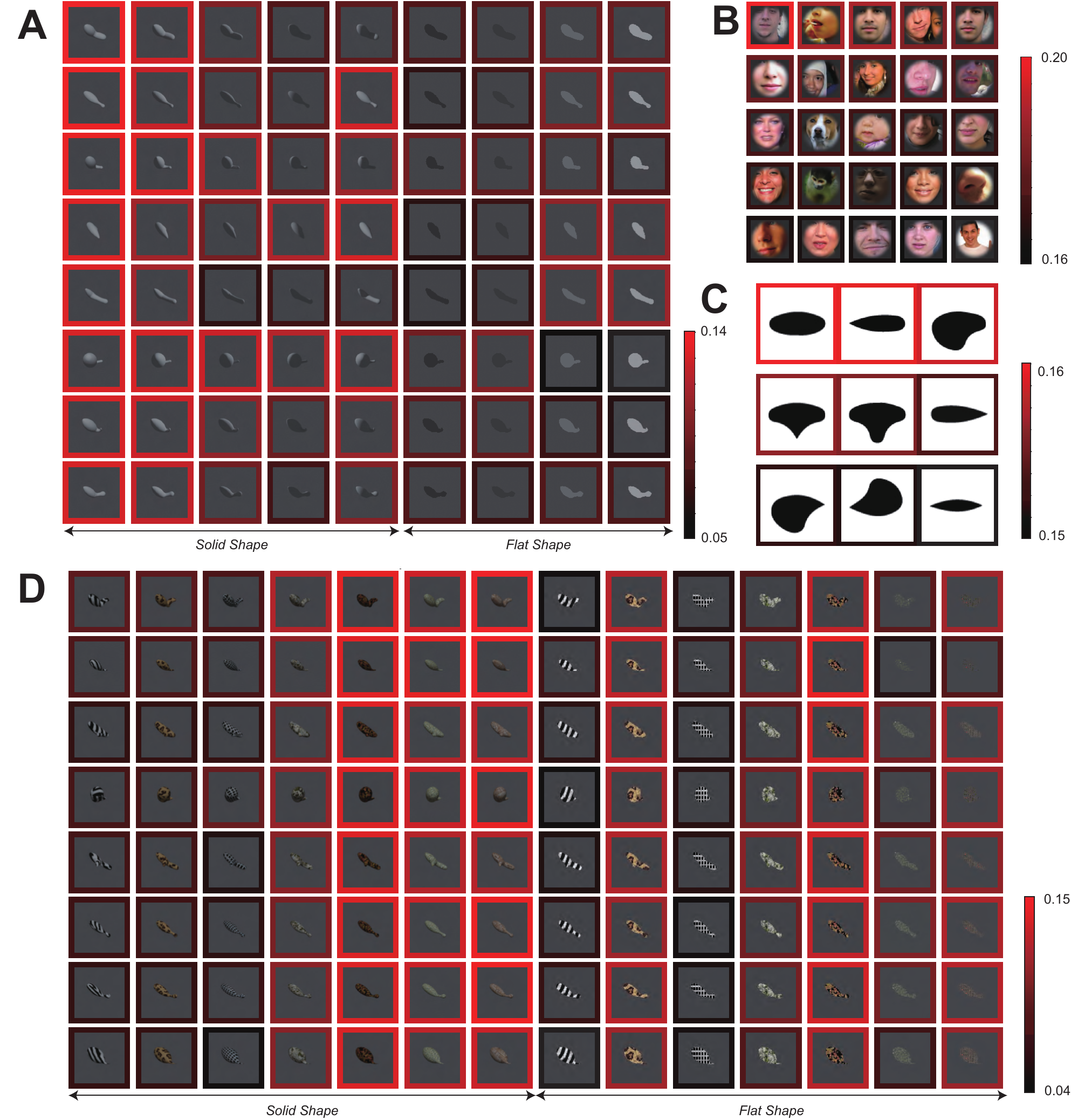}
\end{center}
\caption{\textbf{Area V4 Encodes Solid-Shape Information—Illustrative Examples.}
    \textbf{(A)} Shading stimulus. Normalized response range: 0.05–0.14. Mean Solid–Flat Index (SFI) across lighting directions: 0.0822.
    \textbf{(B)} Preferred natural images. Response range: 0.18–0.20.
    \textbf{(C)} Preferred stimuli from the 2D shape and texture set of \citet{KimBairPasupathy2019a}. Response range: 0.15-0.16.
    \textbf{(D)} \textbf{Texture render.} Normalized response range: $0.04$--$0.15$. Mean SFI across seven textures: $0.3840$.}

  \label{fig:Neuron_2_example}
\end{figure*}

Using the algorithm of \cite{HungCarlsonConnor2012a}, we rendered solid stimuli in Blender and extracted their corresponding flat shapes. We then presented a large set of stimuli to the deep learning model to predict V4 responses. In total, there are 3048 pixels (each pixel occupies $100 \mu \text{m} \times 100 \mu \text{m}$, with about 100 neurons); here we show two illustrative examples.

The first example neuron exhibits a combined texture preference and flat-shape preference. Under the shading paradigm (Fig.\ref{fig:Neuron_1_example}A), responses concentrate on the flat side with a normalized range of $0.02$--$0.10$, yielding a mean SFI across lighting directions of $-0.7752$, consistent with a bias toward flat rather than solid shape. For natural images (Fig.\ref{fig:Neuron_1_example}B), the neuron responds more strongly to texture-rich pictures (range $0.30$--$0.42$). In the 2D shape/texture set of \citet{KimBairPasupathy2019a} (Fig.\ref{fig:Neuron_1_example}C), preferred stimuli again carry prominent texture cues (range $0.37$--$0.43$). In the texture-render condition (Fig.\ref{fig:Neuron_1_example}D), this pattern is reinforced: the normalized response remains $0.02$--$0.10$ and the mean SFI across seven textures is $-0.3930$, indicating that, even with outline and lighting controlled, adding texture systematically elevates responses to flat stimuli. Notably, when probed with a simple uniform–shading stimulus (Fig.~\ref{fig:Neuron_1_example}A), the response occupies a compressed dynamic range (normalized \(0\)–\(0.10\)), markedly lower than the neuron's preferred responses to natural images (\(0.30\)–\(0.42\); Fig.~\ref{fig:Neuron_1_example}B) and to the texture entries of the 2D shape/texture set (\(0.37\)–\(0.43\); Fig.~\ref{fig:Neuron_1_example}C). This indicates that reliance on a single uniform–shading probe (Fig.~\ref{fig:Neuron_1_example}A) might not establish a genuine “flat” preference for this neuron: its responses are markedly stronger when surface statistical content is present (natural images, Fig.~\ref{fig:Neuron_1_example}B; texture entries and texture renders, Fig.~\ref{fig:Neuron_1_example}C, D) than for a plain, uniformly shaded surface.

The second example neuron shows the complementary pattern: a solid-shape preference with relatively weak selectivity to surface texture. In the shading paradigm (Fig.\ref{fig:Neuron_2_example}A), responses span $0.05$--$0.14$ and the mean SFI across lighting directions is positive ($+0.0822$), indicating a bias toward solid (shaded) over flat shapes. For natural images (Fig.\ref{fig:Neuron_2_example}B), preferred exemplars fall in a narrow, higher range ($0.18$--$0.20$), consistent with stronger drive by stimuli containing volumetric shading cues. In the 2D shape/texture set (Fig.\ref{fig:Neuron_2_example}C), responses are moderate and tightly clustered ($0.15$--$0.16$), suggesting limited tuning to outline alone. Under the texture-render condition (Fig.\ref{fig:Neuron_2_example} D), the bias toward solid is amplified: responses remain $0.04$--$0.15$ with a mean SFI of $+0.3840$ across seven textures, showing that adding texture does not overturn the preference for solid. Comparing to the previous neuron, this neuron's shading–evoked response amplitudes fall within the same dynamic range as those observed under the other controlled paradigms—outline-only (Fig.~\ref{fig:Neuron_2_example}C) and texture-render (Fig.~\ref{fig:Neuron_2_example}D)—and are only modestly lower than for natural images (Fig.~\ref{fig:Neuron_2_example}B; \(0.18\)–\(0.20\)). This comparability indicates that the shading manipulation alone provides an adequate and reliable readout of the unit’s preference. Consistently, the condition-matched SFI remains positive when averaged across illumination directions (\(+0.0822\)) and across texture families (\(+0.3840\)), suggesting that the selectivity primarily derives from the solid--flat attribute (3D relief) rather than sensitivity to surface texture per se.

These two exemplars bracket the texture–shape continuum and illustrate how combining the sign of solid preference (via SFI) with a binary texture-preference criterion yields stable single-unit assignments across paradigms (Fig~\ref{fig:Neuron_1_example}, Fig~\ref{fig:Neuron_2_example} panels A–D). In the following, we quantify the population frequency of each category and assess stability when conditions are considered separately (by illumination direction or by texture family).

\subsection{Topological Organization of 3D Solid versus 2D Flat Preferences}

\begin{figure*}[!t]
\begin{center}
\includegraphics[width=\linewidth]{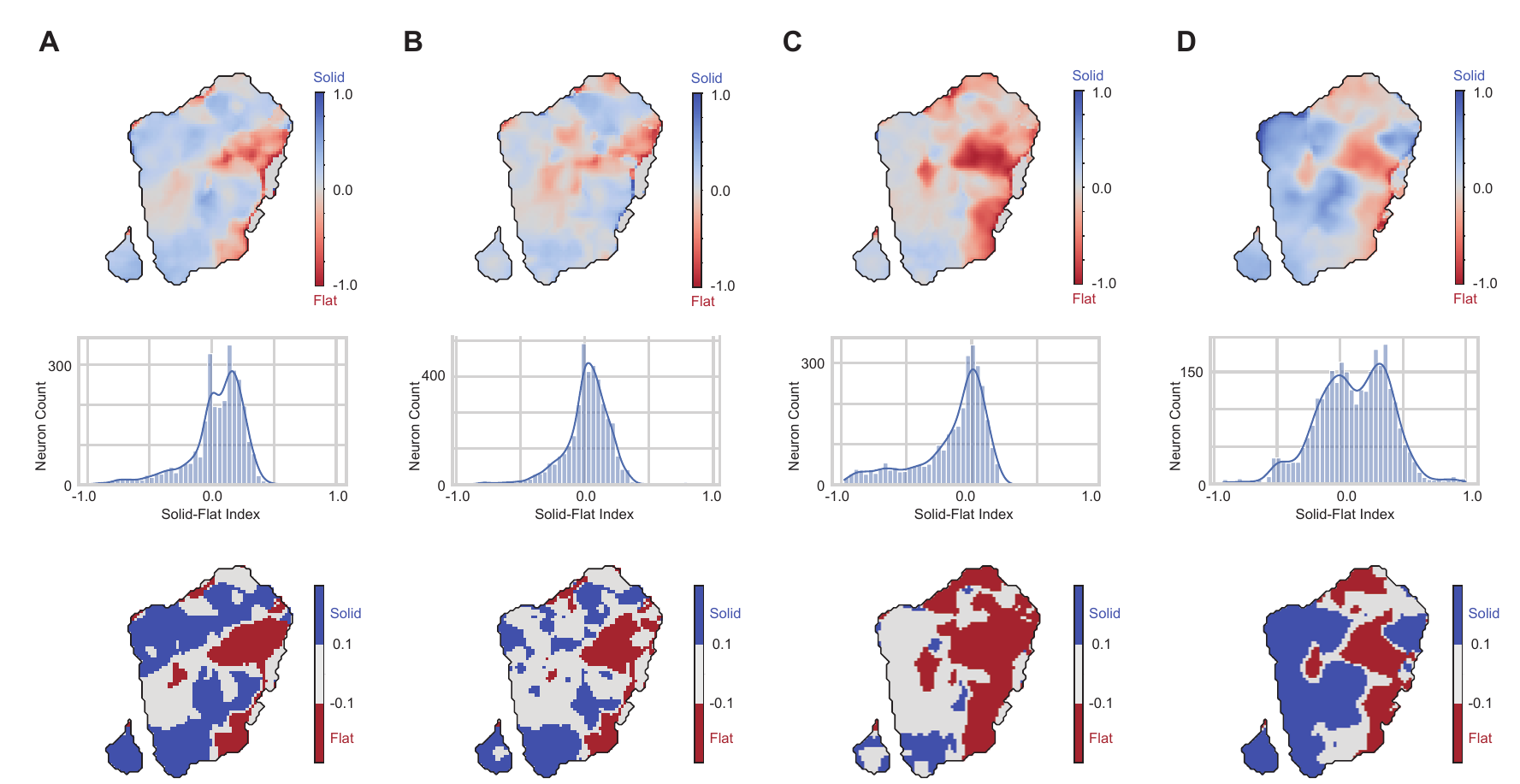}
\end{center}
\caption{
    \textbf{Topographic organization of the Solid--Flat Index (SFI) across stimulus conditions.}
    \textbf{(A)} Shaded Solid/Flat Preference based on one illumination direction: SFI map and corresponding distribution of SFI values for the stimulus set used in \cite{SrinathEmondsWang2021a} with shaded 3D solid rendered based on one direction only.
    \textbf{(B)} Shaded Solid/Flat Preference by integrating results from five illumination directions: SFI map by combining results by averaging five SFI maps, each is based on shading rendered from one of the five illumination directions.  
    \textbf{(C)} Shaded Solid/Flat Preference with 2D control reinforced by noise texture. 
    \textbf{(D)} Shaded and Textured Solid/Flat Preference, averaged across the SFI maps based on five distinct classes of textures. 
    SFI ranges from $-1$ to $1$; negative values indicate a preference for flat stimuli, whereas positive values indicate a preference for solid (shaded) stimuli.
    Second row: The distribution of the SFI values in the four cases. 
    Third row replots and trinarize the first row of SFI maps, setting high solid preference to blue ($\text{SFI}\geq0.1$), high flat preference to red ($\text{SFI}\leq-0.1$), and ambiguous solid/flat preference gray zone ($\text{SFI}$ within the range of -0.1 to 0.1).  
}

  \label{fig:SFI_Map_Distribution}
\end{figure*}

We computed the SFI to quantify the selectivity to solid and flat surface under four stimulus regimes (Eq.~\ref{eq:sfi_ed}, Eq.~\ref{eq:sfi_dense}, Eq.~\ref{eq:sfi_all}, Eq.~\ref{eq:sfi_texture}). We then projected pixelwise SFI values (ranging from $-1$ to $1$, where negative values indicate flat preference and positive values indicate solid preference) onto the cortical sheet to obtain topographic maps, and plotted the corresponding SFI distributions beneath each map (Fig.~\ref{fig:SFI_Map_Distribution}A--D). 

Across conditions, the SFI maps exhibited coherent spatial domains with like preferences clustering together, indicating a systematic mesoscale organization of solid feature coding in V4. Under the shading condition, consistent with \citet{SrinathEmondsWang2021a}, we observed clear clustering of neuronal preferences (Fig.~\ref{fig:SFI_Map_Distribution}A); however, the accompanying histogram shows a broader negative tail (SFI$<0$, flat-preferring), and—unlike the approximately normal distribution reported by \citet{SrinathEmondsWang2021a}—our distribution departs from normality, with evident skew toward negative values. Denser illumination sampling (Fig.~\ref{fig:SFI_Map_Distribution}B) sharpened domain boundaries and increased map contrast, in line with reduced measurement noise arising from lighting variability. Fig.~\ref{fig:SFI_Map_Distribution}C–D summarize average SFI maps. The within-illumination average (Fig.~\ref{fig:SFI_Map_Distribution}C), obtained by pooling across control lighting directions, attenuates illumination confounds and accentuates shape-driven selectivity, yielding a topology that mirrors the full analysis but with cleaner boundaries. The texture-control average (Fig.~\ref{fig:SFI_Map_Distribution}D) shows that although surface texture modulates response amplitudes, the large-scale solid--flat layout is preserved. 

In practice, we observed that neurons with SFI values near zero, $\mathrm{SFI}\in[-0.1,\,0.1]$, can flip polarity when the stimulus condition switches from \emph{shading} to \emph{texture}. To minimize the influence of these borderline cases and to more clearly visualize the topology of neurons preferring solid versus flat, we plot the spatial distribution of three categories: $\mathrm{SFI}>0.1$ (solid-preferring;blue), $\mathrm{SFI}<-0.1$ (flat-preferring; red), and $|\mathrm{SFI}|\le 0.1$ (intermediate; gray), as shown in Fig.~\ref{fig:SFI_Map_Distribution}H. We found that the three-category SFI maps derived from texture renders (right) exhibit strong topographic similarity to the texture-preference map (Fig.~\ref{fig:texture_shape_roi}D) reported above.

To directly compare solid feature coding with the TPI, we computed topological correlations between the SFI map and the TPI map for the first two stimulus sets. For the set following \cite{KimBairPasupathy2019a}, Pearson’s $r=-0.3451$ and Spearman’s $\rho=-0.2843$; with denser illumination sampling, Pearson’s $r=-0.3940$ and Spearman’s $\rho=-0.2939$.

For all shading and texture conditions, to enable a fair comparison with the shape–texture index, we therefore computed topological correlations separately for each shading direction and each texture control. The corresponding condition-specific SFI maps are shown in Fig.~\ref{fig:appendix_condition_SFI_map}. The shading-matched results are reported in Table~\ref{tab:shade_corr_simple}, and the texture-matched results in Table~\ref{tab:text_corr_simple}, each listing both Pearson’s $r$ and Spearman’s $\rho$.Across shading conditions, the mean Pearson correlation was $r=-0.5852\pm0.0483$ and the mean Spearman rank correlation was $\rho=-0.4600\pm0.0900$, whereas across texture conditions, the mean Pearson correlation was $r=-0.6504\pm0.0806$ and the mean Spearman rank correlation was $\rho=-0.6793\pm0.0711$. 

These analyses consistently yielded negative correlations between the TPI and SFI across all lighting directions and texture families. Under the sign convention used here (larger SFI indicates stronger responses to solid relative to flat), a negative correlation implies that pixels preferring 2D shape (higher shape--texture index) tend to show higher SFI (respond more to flat than solid), whereas pixels preferring texture tend to show lower SFI (respond more to solid than flat). In short, 2D shape aligns with solid, whereas texture aligns with flat. 

\begin{table*}[!ht]
\begin{center}
\caption{Correlation between the shape--texture map and SFI computed per shading direction.}
\label{tab:shade_corr_simple}
\vskip 0.12in
\begin{tabular}{lcccccc}
\hline
Metric & Shading 1 & Shading 2 & Shading 3 & Shading 4 & Shading 5 & Average \\
\hline
Pearson  & $-0.5330$ & $-0.5552$ & $-0.6448$ & $-0.6463$ & $-0.5675$ & $-0.5852 \pm 0.0483$ \\
Spearman & $-0.3586$ & $-0.3964$ & $-0.5526$ & $-0.5537$ & $-0.4365$ & $-0.4600 \pm 0.0900$ \\
\hline
\end{tabular}
\end{center}
\end{table*}

\begin{table*}[!ht]
\begin{center}
\caption{Correlation between the shape--texture map and SFI computed per texture control.}
\label{tab:text_corr_simple}
\vskip 0.12in
\begin{tabular}{lcccccccc}
\hline
Metric & Texture 1 & Texture 2 & Texture 3 & Texture 4 & Texture 5 & Texture 6 & Texture 7 & Average \\
\hline
Pearson  & $-0.5548$ & $-0.5767$ & $-0.6895$ & $-0.6965$ & $-0.5699$ & $-0.7150$ & $-0.7507$ & $-0.6504 \pm 0.0806$ \\
Spearman & $-0.5981$ & $-0.6175$ & $-0.7357$ & $-0.6848$ & $-0.5915$ & $-0.7593$ & $-0.7685$ & $-0.6793 \pm 0.0711$ \\
\hline
\end{tabular}
\end{center}
\end{table*}


\subsection{Preference for Shapes over Texture is Correlated with Preference 3D solids} 

\begin{figure*}[!t]
\begin{center}
\includegraphics[width=0.8\linewidth]{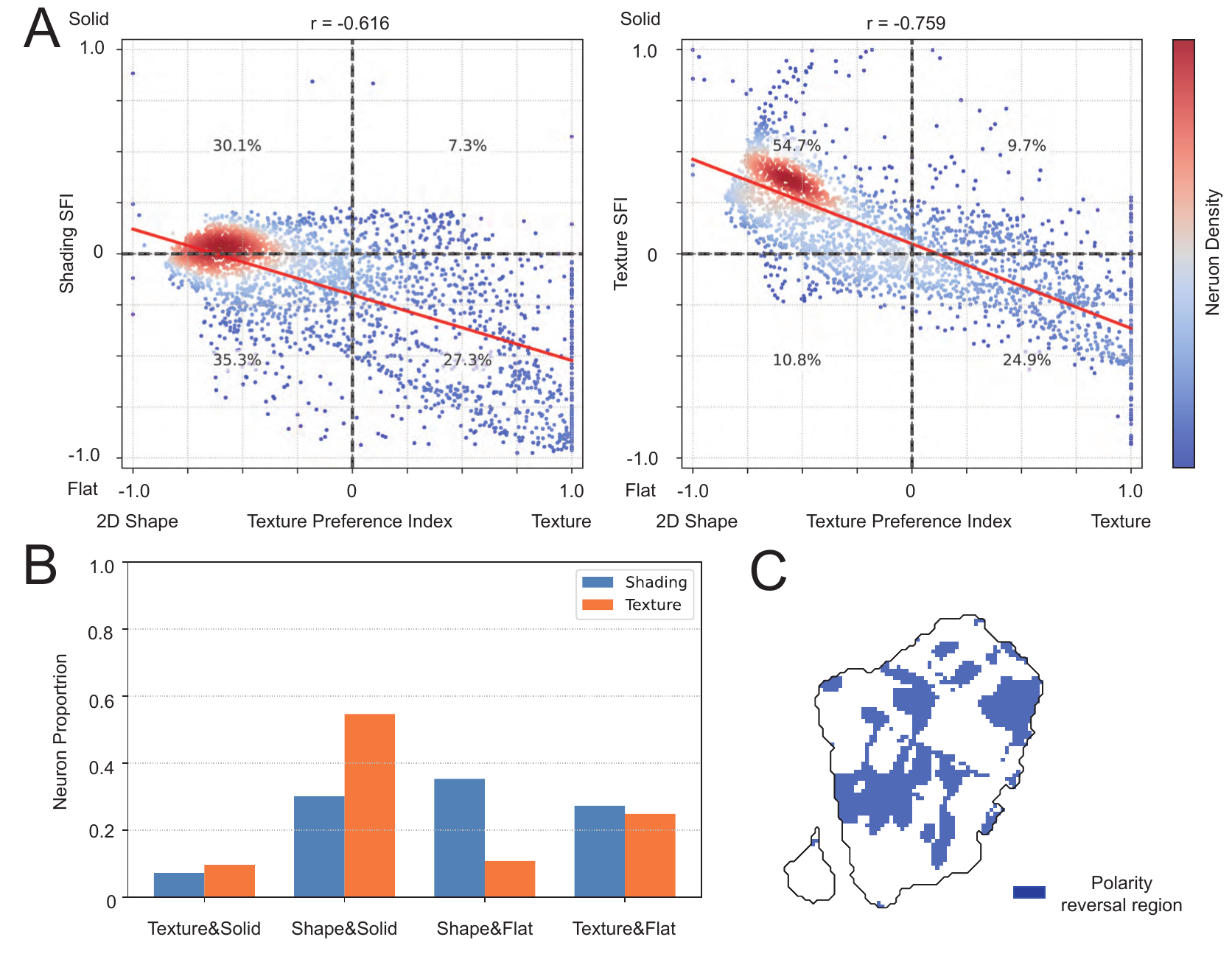}
\end{center}
\caption{\textbf{Coupling between shape–texture preference and solid preference across cue families.}
\textbf{A. Scatter with robust trends.} For each neuron and each matched 3D--2D pair, the SFI is plotted against the neuron's shape–texture preference (X-axis) index using SFI from shading (left) or texture (right) renders (Y-axis). Points are colored by local density; red lines show robust linear fits. Both panels exhibit strong negative correlations (shading $r=-0.568$, texture $r=-0.719$), such that neurons nearer the \emph{shape/curvature} end prefer \emph{solid} (higher SFI), whereas neurons nearer the \emph{texture/dispersed-statistics} end prefer \emph{flat} (lower SFI). 
\textbf{B. Quadrant composition.} Neurons are categorized in the SFI--shape/texture plane: Q1 (Texture--Solid), Q2 (Shape--Solid), Q3 (Shape--Flat), Q4 (Texture--Flat). Bar plots compare shading (orange) and texture (blue). Both cue families are dominated by Q2 and Q4; under texture renders the proportion of Q2 increases while Q3 decreases slightly, with Q4 comparable and Q1 low.
\textbf{C. Polarity-reversal map.} Cortical map of SFI sign flips (blue) from pure shading to adding texture to shading. Flip indicates inconsistent and sites form spatial clusters; in the illustrated case, $787/3048$ neurons ($\approx25.8\%$) change SFI sign, consistent with localized reweighting of cue contributions when texture information is introduced.}
\label{fig:scatter}
\end{figure*}

To get a clear view of why the correlation is negative, for each neuron and each matched 3D--2D stimulus pair, we plotted the SFI against the neuron's shape–texture preference index. As shown in Fig.~\ref{fig:scatter}A, for each neuron and each matched 3D--2D stimulus pair the SFI against the neuron's TPI, using SFI derived from shading (left) and texture (right) renders. Points are colored by local density and red lines denote robust linear fits. Both panels exhibit strong negative correlations (shading $r=-0.616$, texture $r=-0.759$; $N$ as indicated), such that neurons nearer the shape end tend to prefer solid stimuli (higher SFI), whereas neurons nearer the texture end tend to prefer flat stimuli (lower SFI). The steeper slope under texture renders indicates tighter coupling between curvature coding and solid preference when ecologically realistic surface statistics are present.

Consistent with this trend, the quadrant composition in the joint SFI--shape/texture plane (Fig.~\ref{fig:scatter}B) is dominated by Q2 (Shape--Solid) and Q4 (Texture--Flat) across cue families. Relative to shading, adding texture cues  increase the proportion of neurons with Shape--Solid preference (Q2) neurons and reduce neurons with Shape--Flat preference (Q3). Adding texture however does not change the proportions in Q4 (neurons that prefer texture and flat) relative to Q1 (neurons that prefer texture and solid).  This population pattern agrees with the negative slopes in Fig.~\ref{fig:scatter}A and indicates that introducing realistic texture rendering strengthens the consistency between shape curvature coding and solid preference.

To evaluate robustness across cue families, we mapped sites that reverse SFI polarity between shading and texture (Fig.~\ref{fig:scatter}C). Polarity-reversal regions form spatial clusters rather than being uniformly distributed, consistent with localized reweighting of cue contributions. In the illustrated dataset, $787/3048$ neurons ($\approx 25.8\%$) change SFI sign, indicating a substantial minority of locations where solid preference is cue-contingent while the majority preserve a stable mapping across cues.

Taken together, the scatter relationships (Fig.~\ref{fig:scatter}A), the quadrant composition (Fig.~\ref{fig:scatter}B), and the polarity-reversal map (Fig.~\ref{fig:scatter}C) converge on a mesoscale principle:solid preference covaries with shape coding, whereas flat preference covaries with texture, with stronger congruence in the presence of ecologically realistic textures.

Additionally, we grouped neurons by their SFI and texture–preference index across shading and texture conditions, and visualized the joint categories using an additive color–coding scheme. Concretely, for each neuron we binarized two decisions: (i) texture vs.\ 2D–shape preference, encoded as a red channel \(R\) with \(R=1\) if the texture index \(>0.5\) and \(R=0\) otherwise; and (ii) solid vs.\ flat preference, encoded as a green channel \(G\) with \(G=1\) if \(\mathrm{SFI}>0\) and \(G=0\) otherwise. Combining these channels yields four mutually exclusive colors that correspond to the quadrants in Fig.~\ref{fig:scatter}B: red \((R=1,\,G=0)\) indicates texture–preferring with flat preference (\(\mathrm{SFI}<0\))—\emph{consistent}; green \((R=0,\,G=1)\) indicates 2D–shape–preferring with solid preference (\(\mathrm{SFI}>0\))—\emph{consistent}; yellow \((R=1,\,G=1)\) indicates texture–preferring with solid preference—\emph{inconsistent}; and black \((R=0,\,G=0)\) indicates 2D–shape–preferring with flat preference—\emph{inconsistent}. 
Panels \textbf{A} and \textbf{B} in Fig.~\ref{fig:appendix_consistent_map} summarize these consistency maps across shading directions and texture families, respectively.


\subsection{Object Form Selectivity is Correlated with Solid preference }

\begin{figure*}[!h]
\begin{center}
\includegraphics[width=0.8\linewidth]{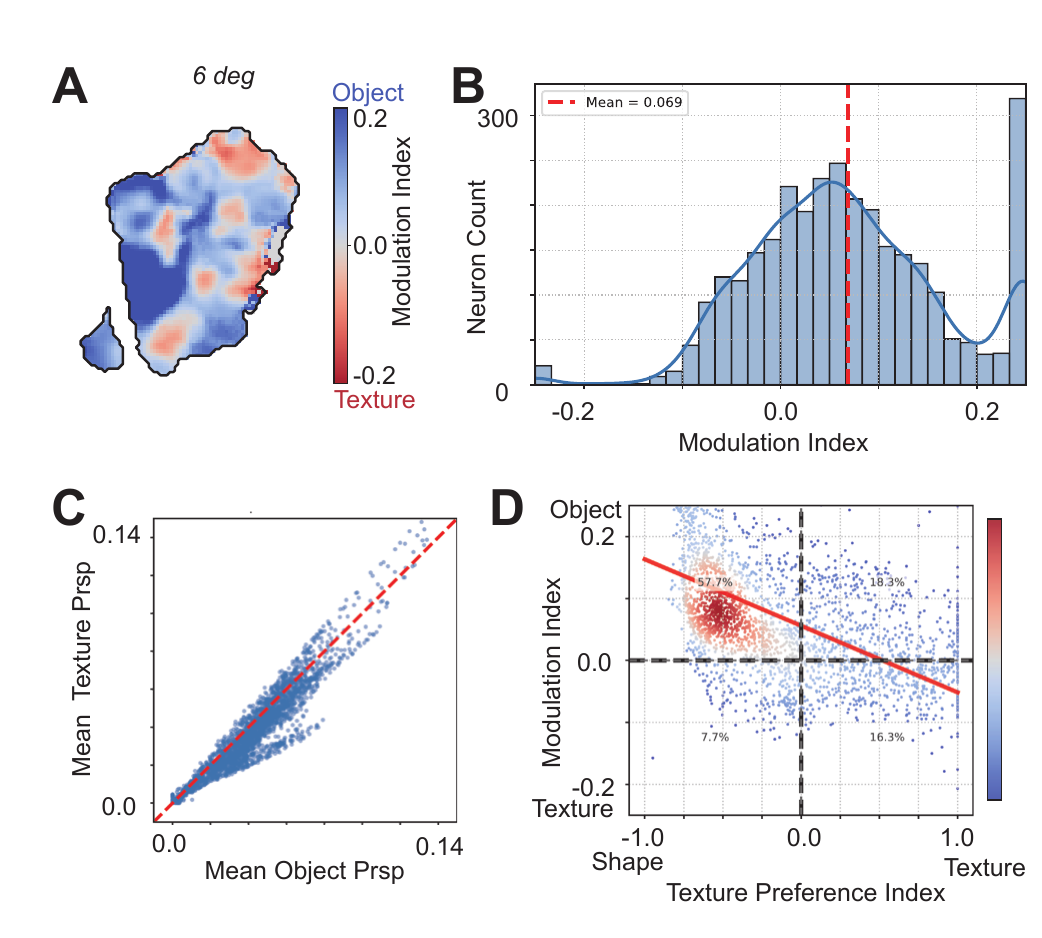}
\end{center}
 \caption{\textbf{Object–Texture Sensitivity.}
  \textbf{(A)} Topological map of the object–texture modulation index (MI; positive = object-dominant, negative = texture-dominant).
  \textbf{(B)} Histogram of MI across neurons.
  \textbf{(C)} Average predicted responses of the V4 digital-twin model to the original object images and their corresponding P--S texture surrogates. The red diagonal line (\(y = x\)) marks equal responses to object and P--S texture stimuli; neurons lying near this line respond similarly to the original objects and their corresponding texture surrogates.
  \textbf{(D)} MI as a function of each neuron’s texture preference index (TPI), with a fitted regression line; vertical dashed lines indicate reference preference levels. The association is negative (Pearson $r = -0.4187$, Spearman $\rho = -0.6139$).
  }
\label{fig:tex_obj_mi}
\end{figure*}

\begin{table*}[!ht]
\begin{center}
\caption{Correlation between the Object--Texture MI map and SFI computed per shading direction.}
\label{tab:tex_obj_shading_corr}
\vskip 0.12in
\begin{tabular}{lcccccc}
\hline
Metric & Shading 1 & Shading 2 & Shading 3 & Shading 4 & Shading 5 & Average \\
\hline
Pearson  & $0.0861$ & $0.0994$ & $0.2053$ & $0.1994$ & $0.1166$ & $0.1414 \pm 0.0508$ \\
Spearman & $0.0219$ & $0.0793$ & $0.2901$ & $0.2685$ & $0.1122$ & $0.1544 \pm 0.1062$ \\
\hline
\end{tabular}
\end{center}
\end{table*}

\begin{table*}[!ht]
\begin{center}
\caption{Correlation between the Object-Texture MI map and SFI computed per texture control.}
\label{tab:text_corr_obj_simple}
\vskip 0.12in
\begin{tabular}{lcccccccc}
\hline
Metric & Texture 1 & Texture 2 & Texture 3 & Texture 4 & Texture 5 & Texture 6 & Texture 7 & Average \\
\hline
Pearson  & $0.2983$ & $0.2708$ & $0.2694$ & $0.2370$ & $0.1612$ & $0.4097$ & $0.3894$ & $0.2909 \pm 0.0797$ \\
Spearman & $0.4028$ & $0.4162$ & $0.4591$ & $0.3751$ & $0.3046$ & $0.5047$ & $0.4491$ & $0.4159 \pm 0.0599$ \\
\hline
\end{tabular}
\end{center}
\end{table*}


The 3D solids and 2D shapes used in Pasupathy and Connor’s experiments are synthetic stimuli \cite{KimBairPasupathy2019a, SrinathEmondsWang2021a}. In natural scenes, however, 3D structure and 2D shape are typically conveyed through object surface structures and object boundaries, rather than isolated geometric primitives. We therefore asked whether selectivity for 3D solids and 2D shapes in our experiments is related to neuronal selectivity for object form in real-world images.

\citet{Lieber2024} investigated V4 neurons’ sensitivity along a form–texture axis, comparing responses to images of intact objects and that to matched scramble texture surrogates generated using the Portilla–Simoncelli (P–S) texture model \cite{Portilla2000}. These texture surrogates preserve local multi-scale, multi-orientation statistics while substantially disrupting global object structure and contour information.  Following their paradigm, we constructed a continuum of stimuli along the form–texture axis for 500 natural images containing distinct objects, enabling us to test how 3D solid and 2D shape selectivity relates to responses to object form in natural scenes.

We present stimuli in Fig.~\ref{fig:stimuli design}H to the digital twins and use Eq.~\ref{eq:mi_tex_obj} to calculate the neurons' sensitivity to texture and objects. For the fully scrambled object condition ($6^{\circ}$), the object–texture MI map exhibited patchy, interleaved domains across the imaged field of view, with both object-dominant ($\mathrm{MI}>0$) and texture-dominant ($\mathrm{MI}<0$) subregions (Fig.~\ref{fig:tex_obj_mi}A). Comparable patchy, interleaved organization was observed for other viewing angles and scrambling scales as well (see Fig.~\ref{fig:appendix_tex_obj}).
The population distribution of single-neuron MI was narrowly centered near zero with a slight skew toward object dominance (Fig.~\ref{fig:tex_obj_mi}B). Only a small fraction of neurons exhibited $|\mathrm{MI}|>0.5$; for readability, the color map in Fig.~\ref{fig:tex_obj_mi} is clipped to $[-0.2,\,0.2]$, which does not alter the qualitative conclusions. For reference, we plotted the V4 digital-twin’s average predicted responses to the original object images and their corresponding P--S texture surrogates (Fig.~\ref{fig:tex_obj_mi}C). The response ratio ($R_{\text{obj}} / R_{\text{tex}}$) for almost all neurons lies near 1, indicating that the P--S model–induced texture stimuli elicit responses that are nearly indistinguishable from those evoked by the original objects. However, a small group of neurons' ratio seams to lower than one, indicating a preference for objects.

We also examined how the object--texture MI map relates to each neuron's shape-texture preference and solid-flat selectivity. First, we quantified the relationship between MI and the TPI at the single-neuron level (Fig.~\ref{fig:tex_obj_mi}D). MI as a function of each neuron's TPI showed a clear negative trend with a fitted regression line; vertical dashed lines indicate reference preference levels. The association is negative (Pearson $r = -0.4167$, Spearman $\rho = -0.6139$), indicating that strongly texture-preferring neurons tend to be more texture-dominant on the object--texture axis.

Second, we correlated the object--texture MI map with SFI maps computed per shading direction and per texture control. The correlation between the object--texture MI map and the SFI map, evaluated using stimuli analogous to those of \citet{SrinathEmondsWang2021a}, was weakly negative (Pearson $r = -0.0456$, Spearman $\rho = -0.1328$).
Correlations with SFI for each shading direction are summarized in Table~\ref{tab:tex_obj_shading_corr}, whereas correlations with SFI for each texture control are reported in Table~\ref{tab:text_corr_obj_simple}. In both analyses, we observed consistent positive correlations of moderate magnitude (average Pearson $r \approx 0.14$ and $0.29$, average Spearman $\rho \approx 0.15$ and $0.41$, respectively), linking object dominance along the object--texture axis to solid-preferring responses (and conversely, texture dominance to flat-preferring responses) in the shading and texture-control conditions. 

The relatively high correlation between the shape--texture preference index and the object--texture modulation index indicates a potential common coding axis along which contour-defined shapes and intact objects lie at one end, and texture-like structure---whether natural texture fills or P-S generated texture surrogates---lie at the other. In contrast, the low correlation with the solid--flat index map derived from solid--flat stimuli lacking texture cues suggests that this dimension may reflect a partly distinct coding axis, more specifically tied to 3D surface structure independent of texture.

\section{Discussion}

In this study, we leverage V4 digital twins—deep networks fitted to neuronal population responses in V4—to obtain a more integrative understanding of V4 neural codes. Classical studies have emphasized distinct, sometimes contrasting dimensions of V4 coding, including Pasupathy lab’s Shape vs. Texture \cite{KimBairPasupathy2019a}, Connor lab’s 3D vs. 2D Surface \cite{SrinathEmondsWang2021a}, and Movshon lab’s Form vs. Texture \cite{Lieber2024}. Using the digital twins, we can now map the topological organization of these visual attributes across the cortical surface of V4 and quantify their correspondence.

Building on this framework, we asked whether cortical maps organizing solid–flat preference align with those organizing shape–texture selectivity or object-texture selectivity. Classic single-unit work established that V4 neurons encode moderately complex boundary features (curvature at specific angular positions), furnishing a mid-level scaffold for object shape representation \cite{PasupathyConnor2001a}. Subsequent work showed that V4 neurons are selective for higher-order texture statistics and that texture selectivity strengthens from V2 to V4, indicating a surface-oriented channel in mid-level ventral cortex \cite{Okazawa2015V4TextureStats}. At the mesoscale, our maps link this surface channel to 3D/2D representation by aligning the shape–texture and solid–flat axes: shape-preferring domains preferentially overlap with solid-preferring domains, whereas texture-preferring domains preferentially overlap with flat-preferring domains. Together with classic evidence for contour- and boundary-based shape coding in V4, this domain-wise alignment supports a view of V4 as a midtier hub that integrates contour geometry with surface cues \cite{Roe2012UnifiedV4}.

To interpret these “3D” effects appropriately, it is important to clarify what we mean by “3D” in this paradigm. Both “3D (solid)” and “2D (flat)” conditions are ultimately presented as images: the 3D perception arises from manipulating photometric and textural cues (e.g., shading direction, highlight structure, surface texture) rather than from access to true 3D geometry. Classical shape-from-shading is underconstrained (e.g., the bas-relief family) and relies on idealized reflectance and lighting assumptions \cite{Horn1975SFS,Belhumeur1999BasRelief,Basri2003LambertianSubspace}. Our design explicitly builds on this image-based regime: by sampling multiple illumination directions, adding ecologically realistic textures, and including blob- and noise-based “flat” controls, we make the inferred solid–flat axis more robust to any single shading configuration while strengthening the 2D baseline against which solids is read out. The resulting organization is best interpreted as image-induced 3D perception—lawful interactions between boundary- and surface-based codes operating on 2D images—rather than a direct readout of full geometric structure, and naturally motivates future work incorporating additional depth cues such as disparity, motion parallax, dynamic shadows, and contact geometry.

Two observations support a shared solid–flat axis jointly engaged by shading and texture cues. First, SFI maps derived from all four stimulus regimes (shading, dense illumination sampling, pooled illumination, and adding texture to shading controls) exhibit coherent solid- and flat-preferring domains whose large-scale layout is preserved across changes in illumination direction and surface texture; three-category SFI maps from texture renders further show strong topographic similarity to the texture-preference map, indicating that 3D coding is robust to surface statistics. Second, topological map correlations between the SFI and the shape–texture map are consistently negative across stimulus sets, shading directions, and texture families (with $|r|\approx 0.35$ for pooled maps and $|r|\approx 0.6$–$0.7$ for condition-matched analyses; Tables~\ref{tab:shade_corr_simple}–\ref{tab:text_corr_simple}), such that pixels nearer the 2D-shape end of the shape–texture axis tend to prefer solid whereas pixels nearer the texture end tend to prefer flat. These results dovetail with two-photon evidence for segregated solid- and flat-preferring modules \cite{SrinathEmondsWang2021a} and with electrophysiology showing non-trivial contour–texture integration in V4 \cite{KimBairPasupathy2019a}.

Functionally, the joint geometry of these three indices suggests that V4 implements at least two partially independent coding axes: one distinguishing 3D solids from flat surfaces and another distinguishing intact objects from texture-like structure. Jointly, the SFI and TPI define a two-dimensional coordinate frame over V4: one axis parameterizes solid versus flat responses, the other parameterizes shape/object versus texture responses. Most pixels fall along the “shape+solid / texture+flat” diagonal, whereas a minority occupy the off-diagonal quadrants, indicating partially aligned yet separable cue dimensions. The relatively high correlation between the shape--texture preference index and the object--texture modulation index implies that the shape--texture axis lies obliquely in this space, sharing variance with the object--texture axis such that contour-defined shapes and intact objects cluster toward one end, while both natural texture fills and P--S texture surrogates cluster toward the other. By contrast, the weak correlation between the solid--flat index and this object/texture-related dimension indicates that the solid--flat axis is approximately orthogonal to the object--texture axis, capturing variance that is more specifically tied to 3D surface structure independent of texture. This division of labor is consistent with clustered curvature/corner domains in V4 \cite{Jiang2021V4CurvesCorners} and with texture-statistics encoding work showing that P--S features capture texture-driven responses in human V4 and adjacent ventral regions, highlighting higher-order texture statistics as an explicit axis for surface-based representations \cite{Henderson2023TextureEncoding}. Together, these findings support a view in which curvature-based boundary codes and texture-based surface codes are arranged along partially orthogonal axes, yet both project onto a broader shape--texture dimension to yield cue-reliable 3D versus 2D interpretations of complex natural objects.

A minority of sites (roughly one quarter) exhibit the opposite pairing (e.g., texture+solid or shape+flat), and these reversals cluster spatially. We interpret these “inconsistencies” not as violations of a common solid-flat axis but as evidence for context-dependent selective recruitment of different domain networks. In patches where texture cues are more diagnostic than shading for inferring relief (or vice versa), local networks dominated by texture- or shape-preferring units may be preferentially engaged, producing lawful sign reversals in SFI as cue reliability and task demands vary. This form of flexible, cue- and goal-dependent selection echoes Roe et al.’s domain-based account of V4, in which bottom-up feature evidence and top-down attention selectively recruit distinct functional subnets \cite{Roe2012UnifiedV4}, and is consistent with high-density recordings that reveal sparse clusters of similarly tuned neurons without broad uniform columns—reminding us that mesoscale maps can emerge from finer-scale mosaics and method-dependent sampling \cite{Namimae1893232024}.


A methodological limitation concerns spatial resolution. Our unit of analysis is a mesoscale pixel (hundreds of microns) that pools signals from many neurons with heterogeneous tuning, compressing dynamic range, broadening response distributions, and potentially blurring or even flipping apparent solid/flat polarity near zero when cue configurations change. By contrast, single-unit recordings isolate sharply tuned neurons with clearer shape-texture dissociations \cite{KimBairPasupathy2019a}. Nevertheless, mesoscale and cellular measurements appear broadly consistent: in the 3D/2D study of \citet{SrinathEmondsWang2021a}, two-photon maps reflected underlying single-unit biases \cite{SrinathEmondsWang2021a}, and preliminary cell-resolved analyses in our data \cite{Wang2024} likewise find that most neurons within a pixel share its solid/flat and shape/texture preference. Thus, our pixel-level maps are best viewed as pooled summaries of a heterogeneous micro-organization rather than as contradicting classic single-cell results.

In future work, we aim to identify more principled ways to decompose V4 neural codes into orthogonal dimensions, explore co-activation manifolds of features, and search for new intermediate-complexity patterns or intermediate representations (such as figure-ground, geons, or intrinsic images, scene descriptors) beyond previously defined axes. In this way, digital twins can move us beyond reconciling prior results toward a new, generative discovery framework for understanding the V4 code.

Taken together, our V4 digital twin shows how contrasting attribute maps—shape, texture, 2D/3D, and object—can be placed onto a common topographic framework and related to functional domains of natural-image preference. At the same time, it provides a way to design new experiments that move beyond time-limited, single-unit, single-axis protocols by better controlling confounds in classical stimuli (e.g., Pasupathy’s flat textures bounded by 2D contours \cite{KimBairPasupathy2019a}, or contrast and mean-luminance constraints in Connor’s stimuli\cite{SrinathEmondsWang2021a}). By simulating responses to virtually unlimited banks of controlled synthetic image stimuli and evaluating them with perceptual and image-based metrics before recording, the V4 digital twin becomes a powerful in silico preparation for more comprehensive tests of the neural code.

\section{Conclusion}

In this study, we used V4 digital twins to unify and extend our understanding of V4 neural codes for shape versus texture, 3D solid versus 2D surfaces, and object form. By probing these dimensions within a common framework and projecting them onto a shared topological map, we uncovered systematic relationships across these tuning properties. Specifically, selectivity for 2D contour-defined shape aligns with selectivity for 3D solids and object forms, whereas selectivity for uniform texture and luminance aligns with neural codes for 2D flattened surface appearance. These findings suggest an intriguing possibility that V4 may decompose visual inputs into two complementary mid-level representations: {\em a shape module}, which encodes the 3D geometry of objects, and {\em an appearance module}, which encodes surface properties such as texture and color abstracted away from geometry. These representations are expressed in interleaved functional domains across the V4 topographic map. Proving this hypothesis requires further research. The digital-twin approach explored in this paper might provide a platform for designing new experiments to investigate this hypothesis as well as to probe more comprehensively the multidimensional V4 code.

\bibliographystyle{unsrtnat}   

\bibliography{Arxiv}  

@Article{Wang2024,
    author =	 {T. Wang and T. S. Lee and H. Yao and J. Hong and Y. Li and H. Jiang and I. M. Andolina and S. Tang},
    title =	 {Large-scale calcium imaging reveals a systematic V4 map for encoding natural scenes},
    journal =	 {Nature Communications},
    year =	 2024,
    volume =	 15,
    number =	 1,
    pages =	 {6401},
    doi =         {10.1038/s41467-024-50821-z},
    url = {https://doi.org/10.1038/s41467-024-50821-z},
}

@INPROCEEDINGS{Imagenet,
  author={Deng, Jia and Dong, Wei and Socher, Richard and Li, Li-Jia and Kai Li and Li Fei-Fei},
  booktitle={2009 IEEE Conference on Computer Vision and Pattern Recognition}, 
  title={ImageNet: A large-scale hierarchical image database}, 
  year={2009},
  volume={},
  number={},
  pages={248-255},
  doi={10.1109/CVPR.2009.5206848}
}

@Article{KimBairPasupathy2019a,
    author =	 {T. Kim and W. Bair and A. Pasupathy},
    title =	 {Neural coding for shape and texture in macaque area V4},
    journal =	 {Journal of Neuroscience},
    year =	 2019,
    volume =	 39,
    number =	 24,
    pages =	 {4760--4774},
    doi =	 {10.1523/JNEUROSCI.3073-18.2019},
    url =	 {[https://doi.org/10.1523/JNEUROSCI.3073-18.2019]}
}

@Article{PasupathyConnor2001a,
    author = {A. Pasupathy and C. E. Connor},
    title = {Shape representation in area V4: position-specific tuning for boundary conformation},
    journal = {Journal of Neurophysiology},
    year = 2001,
    volume = 86,
    number = 5,
    pages = {2505--2519},
    doi = {10.1152/jn.2001.86.5.2505},
    url = {https://doi.org/10.1152/jn.2001.86.5.2505}
}

@Book{Brodatz1966a,
    author = {P. Brodatz},
    title = {Textures: A Photographic Album for Artists and Designers},
    publisher = {Dover Publications},
    year = 1966,
    address = {New York}
}

@Article{SrinathEmondsWang2021a,
author =	 {R. Srinath and A. Emonds and Q. Wang and A. A. Lempel and E. Dunn-Weiss and C. E. Connor and K. J. Nielsen},
title =	 {Early Emergence of Solid Shape Coding in Natural and Deep Network Vision},
journal =	 {Current Biology},
year =	 2021,
volume =	 31,
number =	 1,
pages =	 {51--65.e5},
doi =	 {10.1016/j.cub.2020.09.076},
url =	 {https://doi.org/10.1016/j.cub.2020.09.076}
}

@Article{HungCarlsonConnor2012a,
author =	 {C.-C. Hung and E. T. Carlson and C. E. Connor},
title =	 {Medial axis shape coding in macaque inferotemporal cortex},
journal =	 {Neuron},
year =	 2012,
volume =	 74,
number =	 6,
pages =	 {1099--1113},
doi =	 {10.1016/j.neuron.2012.04.029},
url =	 {https://doi.org/10.1016/j.neuron.2012.04.029}
}

@article {Lieber2024,
	author = {Lieber, Justin D. and Oleskiw, Timothy D. and Palmieri, Laura and Simoncelli, Eero P. and Movshon, J. Anthony},
	title = {Responses of neurons in macaque V4 to object and texture images},
	elocation-id = {2024.02.20.581273},
	year = {2025},
	doi = {10.1101/2024.02.20.581273},
	publisher = {Cold Spring Harbor Laboratory},
	URL = {https://www.biorxiv.org/content/early/2025/03/04/2024.02.20.581273},
	eprint = {https://www.biorxiv.org/content/early/2025/03/04/2024.02.20.581273.full.pdf},
	journal = {bioRxiv}
}

@article{
JohannesMehrer2021,
author = {Johannes Mehrer  and Courtney J. Spoerer  and Emer C. Jones  and Nikolaus Kriegeskorte  and Tim C. Kietzmann },
title = {An ecologically motivated image dataset for deep learning yields better models of human vision},
journal = {Proceedings of the National Academy of Sciences},
volume = {118},
number = {8},
pages = {e2011417118},
year = {2021},
doi = {10.1073/pnas.2011417118},
URL = {https://www.pnas.org/doi/abs/10.1073/pnas.2011417118},
eprint = {https://www.pnas.org/doi/pdf/10.1073/pnas.2011417118},
}

@article{Freeman2011,
  author   = {Jeremy Freeman and Eero P. Simoncelli},
  title    = {Metamers of the ventral stream},
  journal  = {Nature Neuroscience},
  year     = {2011},
  volume   = {14},
  number   = {9},
  pages    = {1195--1201},
  date     = {2011-09-01},
  issn     = {1546-1726},
  doi      = {10.1038/nn.2889},
  url      = {https://doi.org/10.1038/nn.2889},
}

@article{Portilla2000,
  author   = {Javier Portilla and Eero P. Simoncelli},
  title    = {A Parametric Texture Model Based on Joint Statistics of Complex Wavelet Coefficients},
  journal  = {International Journal of Computer Vision},
  year     = {2000},
  volume   = {40},
  number   = {1},
  pages    = {49--70},
  date     = {2000-10-01},
  issn     = {1573-1405},
  doi      = {10.1023/A:1026553619983},
  url      = {https://doi.org/10.1023/A:1026553619983},
}

@article{Roe2012UnifiedV4,
  author  = {Roe, Anna W. and Chelazzi, Leonardo and Connor, Charles E. and Conway, Bevil R. and Fujita, Ichiro and Gallant, Jack L. and Lu, Haidong and Vanduffel, Wim},
  title   = {Toward a Unified Theory of Visual Area V4},
  journal = {Neuron},
  year    = {2012},
  volume  = {74},
  number  = {1},
  pages   = {12--29},
  month   = apr,
  publisher = {Elsevier},
  issn    = {0896-6273},
  doi     = {10.1016/j.neuron.2012.03.011},
  url     = {https://doi.org/10.1016/j.neuron.2012.03.011},
  urldate = {2025-11-07}
}

@article{Jiang2021V4CurvesCorners,
  author    = {Jiang, Rundong and Andolina, Ian Max and Li, Ming and Tang, Shiming},
  title     = {Clustered functional domains for curves and corners in cortical area V4},
  journal   = {eLife},
  year      = {2021},
  volume    = {10},
  pages     = {e63798},
  month     = may,
  issn      = {2050-084X},
  doi       = {10.7554/eLife.63798},
  pmid      = {33998459},
  pmcid     = {PMC8175081},
  publisher = {eLife Sciences Publications, Ltd},
  url       = {https://doi.org/10.7554/eLife.63798}
}

@article{Henderson2023TextureEncoding,
  author    = {Henderson, Margaret M. and Tarr, Michael J. and Wehbe, Leila},
  title     = {A Texture Statistics Encoding Model Reveals Hierarchical Feature Selectivity across Human Visual Cortex},
  journal   = {The Journal of Neuroscience},
  year      = {2023},
  volume    = {43},
  number    = {22},
  pages     = {4144--4161},
  month     = may,
  doi       = {10.1523/JNEUROSCI.1822-22.2023},
  pmid      = {37127366},
  pmcid     = {PMC10255092},
  issn      = {0270-6474},
  eissn     = {1529-2401},
  publisher = {Society for Neuroscience},
  url       = {https://doi.org/10.1523/JNEUROSCI.1822-22.2023}
}

@article {Namimae1893232024,
	author = {Namima, Tomoyuki and Kempkes, Erin and Zamarashkina, Polina and Owen, Natalia and Pasupathy, Anitha},
	title = {High-Density Recording Reveals Sparse Clusters (But Not Columns) for Shape and Texture Encoding in Macaque V4},
	volume = {45},
	number = {5},
	elocation-id = {e1893232024},
	year = {2025},
	doi = {10.1523/JNEUROSCI.1893-23.2024},
	publisher = {Society for Neuroscience},
	issn = {0270-6474},
	URL = {https://www.jneurosci.org/content/45/5/e1893232024},
	eprint = {https://www.jneurosci.org/content/45/5/e1893232024.full.pdf},
	journal = {Journal of Neuroscience}
}

@incollection{Horn1975SFS,
  author    = {Horn, Berthold K. P.},
  title     = {Obtaining Shape from Shading Information},
  booktitle = {The Psychology of Computer Vision},
  editor    = {Winston, Patrick Henry},
  year      = {1975},
  publisher = {McGraw-Hill},
  address   = {New York},
  pages     = {115--155}
}

@article{Belhumeur1999BasRelief,
  author  = {Belhumeur, Peter N. and Kriegman, David J. and Yuille, Alan L.},
  title   = {The Bas-Relief Ambiguity},
  journal = {International Journal of Computer Vision},
  year    = {1999},
  volume  = {35},
  number  = {1-2},
  pages   = {33--44},
  doi     = {10.1023/A:1008154927611}
}

@article{Basri2003LambertianSubspace,
  author  = {Basri, Ronen and Jacobs, David W.},
  title   = {Lambertian Reflectance and Linear Subspaces},
  journal = {IEEE Transactions on Pattern Analysis and Machine Intelligence},
  year    = {2003},
  volume  = {25},
  number  = {2},
  pages   = {218--233},
  doi     = {10.1109/TPAMI.2003.1177153}
}

@article{Tanigawa2010V4,
  title   = {Functional organization for color and orientation in macaque {V4}},
  author  = {Tanigawa, Hisashi and Lu, Haidong D. and Roe, Anna W.},
  journal = {Nature Neuroscience},
  year    = {2010},
  volume  = {13},
  number  = {12},
  pages   = {1542--1548},
  doi     = {10.1038/nn.2676},
  pmid    = {21076422},
  pmcid   = {PMC3005205},
  issn    = {1097-6256},
  eissn   = {1546-1726},
  month   = dec,
  note    = {Epub 2010 Nov 14}
}

@article{Arcizet2009,
author = {Arcizet, Fabrice and Jouffrais, Christophe and Girard, Pascal},
year = {2009},
month = {11},
pages = {140},
title = {Coding of shape from shading in area V4 of the macaque monkey},
volume = {10},
journal = {BMC neuroscience},
doi = {10.1186/1471-2202-10-140}
}

@article {Tang2020V4,
article_type = {journal},
title = {Curvature-processing domains in primate V4},
author = {Tang, Rendong and Song, Qianling and Li, Ying and Zhang, Rui and Cai, Xingya and Lu, Haidong D},
editor = {Krug, Kristine and Gold, Joshua I and Connor, Ed and Das, Aniruddha},
volume = 9,
year = 2020,
month = {nov},
pub_date = {2020-11-19},
pages = {e57502},
citation = {eLife 2020;9:e57502},
doi = {10.7554/eLife.57502},
url = {https://doi.org/10.7554/eLife.57502},
journal = {eLife},
issn = {2050-084X},
publisher = {eLife Sciences Publications, Ltd},
}

@article{NANDY20131102,
title = {The Fine Structure of Shape Tuning in Area V4},
journal = {Neuron},
volume = {78},
number = {6},
pages = {1102-1115},
year = {2013},
issn = {0896-6273},
doi = {https://doi.org/10.1016/j.neuron.2013.04.016},
url = {https://www.sciencedirect.com/science/article/pii/S0896627313003188},
author = {Anirvan S. Nandy and Tatyana O. Sharpee and John H. Reynolds and Jude F. Mitchell},
}

@article {Li202V4,
	author = {Li, Ming and Liu, Fang and Juusola, Mikko and Tang, Shiming},
	title = {Perceptual Color Map in Macaque Visual Area V4},
	volume = {34},
	number = {1},
	pages = {202--217},
	year = {2014},
	doi = {10.1523/JNEUROSCI.4549-12.2014},
	publisher = {Society for Neuroscience},
	issn = {0270-6474},
	URL = {https://www.jneurosci.org/content/34/1/202},
	eprint = {https://www.jneurosci.org/content/34/1/202.full.pdf},
	journal = {Journal of Neuroscience}
}

@article{Kobatake1994ComplexObject,
  title   = {Neuronal selectivities to complex object features in the ventral visual pathway of the macaque cerebral cortex},
  author  = {Kobatake, E. and Tanaka, K.},
  journal = {Journal of Neurophysiology},
  year    = {1994},
  volume  = {71},
  number  = {3},
  pages   = {856--867},
  month   = mar,
  doi     = {10.1152/jn.1994.71.3.856},
  issn    = {0022-3077},
  pmid    = {8201425}
}

@article{Henry2022CrowdingV1V4,
  title   = {Feature representation under crowding in macaque V1 and V4 neuronal populations},
  author  = {Henry, Christopher A. and Kohn, Adam},
  journal = {Current Biology},
  year    = {2022},
  volume  = {32},
  number  = {23},
  pages   = {5126--5137.e3},
  month   = dec,
  doi     = {10.1016/j.cub.2022.10.049},
  pmid    = {36379216},
  pmcid   = {PMC9729449},
  issn    = {0960-9822},
  eissn   = {1879-0445},
  note    = {Epub 2022 Nov 14}
}

@article{Smith2020,
author = "Matthew Smith",
title = "{V4 Utah Array Plaid Movie Data}",
year = "2020",
month = "5",
url = "https://figshare.com/articles/dataset/V4_Utah_Array_Plaid_Movie_Data/12269513",
doi = "10.6084/m9.figshare.12269513.v1"
}

@article{Okazawa2015V4TextureStats,
  title   = {Image statistics underlying natural texture selectivity of neurons in macaque V4},
  author  = {Okazawa, Gouki and Tajima, Satohiro and Komatsu, Hidehiko},
  journal = {Proceedings of the National Academy of Sciences of the United States of America},
  year    = {2015},
  volume  = {112},
  number  = {4},
  pages   = {E351--E360},
  month   = jan,
  doi     = {10.1073/pnas.1415146112},
  pmid    = {25535362},
  pmcid   = {PMC4313822},
  issn    = {0027-8424},
  eissn   = {1091-6490},
  note    = {Epub 2014 Dec 22}
}

@article{Hu2020CurvatureV4,
  title   = {Curvature domains in V4 of macaque monkey},
  author  = {Hu, Jia Ming and Song, Xue Mei and Wang, Qiannan and Roe, Anna Wang},
  journal = {eLife},
  year    = {2020},
  volume  = {9},
  pages   = {e57261},
  month   = nov,
  doi     = {10.7554/eLife.57261},
  pmid    = {33211004},
  pmcid   = {PMC7707819},
  issn    = {2050-084X},
  note    = {Epub 2020 Nov 19}
}

@article {Zhang2023SFV4,
article_type = {journal},
title = {Spatial frequency representation in V2 and V4 of macaque monkey},
author = {Zhang, Ying and Schriver, Kenneth E and Hu, Jia Ming and Roe, Anna Wang},
editor = {Ray, Supratim and Moore, Tirin and Connor, Charles E},
volume = 12,
year = 2023,
month = {jan},
pub_date = {2023-01-06},
pages = {e81794},
citation = {eLife 2023;12:e81794},
doi = {10.7554/eLife.81794},
url = {https://doi.org/10.7554/eLife.81794},
journal = {eLife},
issn = {2050-084X},
publisher = {eLife Sciences Publications, Ltd},
}

@article{Arcizet2008NaturalTextures,
  author  = {Arcizet, F. and Jouffrais, C. and Girard, P.},
  title   = {Natural textures classification in area V4 of the macaque monkey},
  journal = {Experimental Brain Research},
  year    = {2008},
  volume  = {189},
  number  = {1},
  pages   = {109--120},
  doi     = {10.1007/s00221-008-1406-9},
  issn    = {0014-4819},
  pmid    = {18506435}
}

@article{MERIGAN_2000, title={Cortical area V4 is critical for certain texture discriminations, but this effect is not dependent on attention}, volume={17}, DOI={10.1017/S095252380017614X}, number={6}, journal={Visual Neuroscience}, author={Merigan, William H.}, year={2000}, pages={949–958}}

@article{WALSH199351,
title = {The effects of V4 lesions on the visual abilities of macaques: hue discrimination and colour constancy},
journal = {Behavioural Brain Research},
volume = {53},
number = {1},
pages = {51-62},
year = {1993},
issn = {0166-4328},
doi = {https://doi.org/10.1016/S0166-4328(05)80265-7},
url = {https://www.sciencedirect.com/science/article/pii/S0166432805802657},
author = {V. Walsh and D. Carden and S.R. Butler and J.J. Kulikowski},

}

@article{PASUPATHY2019199,
title = {Object shape and surface properties are jointly encoded in mid-level ventral visual cortex},
journal = {Current Opinion in Neurobiology},
volume = {58},
pages = {199-208},
year = {2019},
note = {Computational Neuroscience},
issn = {0959-4388},
doi = {https://doi.org/10.1016/j.conb.2019.09.009},
url = {https://www.sciencedirect.com/science/article/pii/S0959438818302654},
author = {Anitha Pasupathy and Taekjun Kim and Dina V Popovkina},
}

\appendix

\section{Supplementary Figures}
\begin{figure*}
    \centering
    \includegraphics[width=\linewidth]{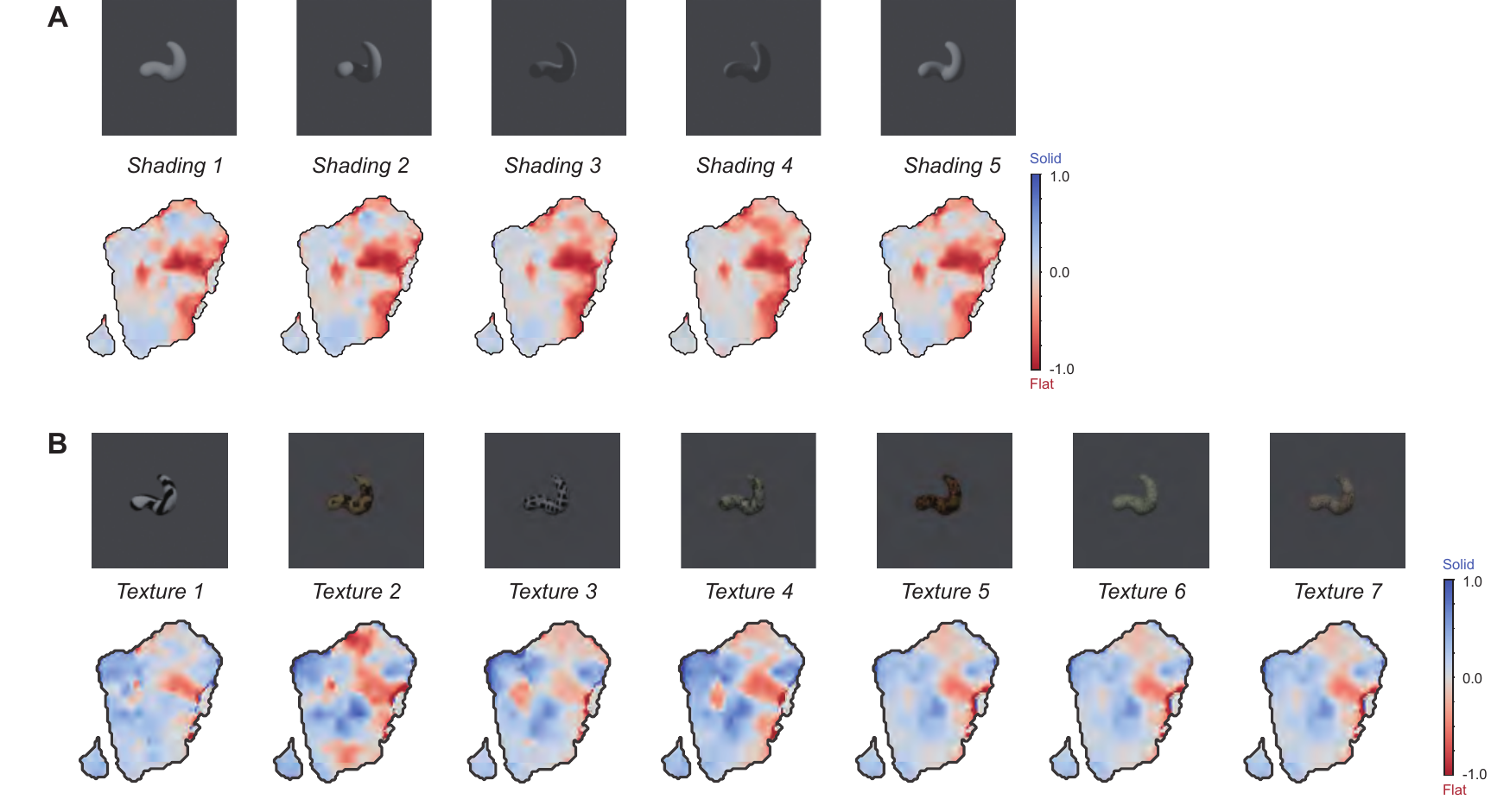}
    \caption{Solid–flat index (SFI) maps summarizing solid-flat preference within each illumination condition and across texture-control stimuli. 
    \textbf{(A)} SFI maps for five lighting directions, with insets illustrating example solid and flat objects under each lighting direction.
    \textbf{(B)} SFI maps for seven textures added to the shading stimuli, with insets illustrating example shaded and texture-plus-shading objects for each texture condition.
    SFI values range from $-1$ to $1$. Positive values (blue; solid-preferring) indicate stronger responses to solid than flat stimuli, whereas negative values (red; flat-preferring) indicate stronger responses to flat than solid stimuli. Example solid and flat objects for each condition are shown alongside the corresponding maps.}
    \label{fig:appendix_condition_SFI_map}
\end{figure*}

\begin{figure*}[!t]
    \centering
    \includegraphics[width=\linewidth]{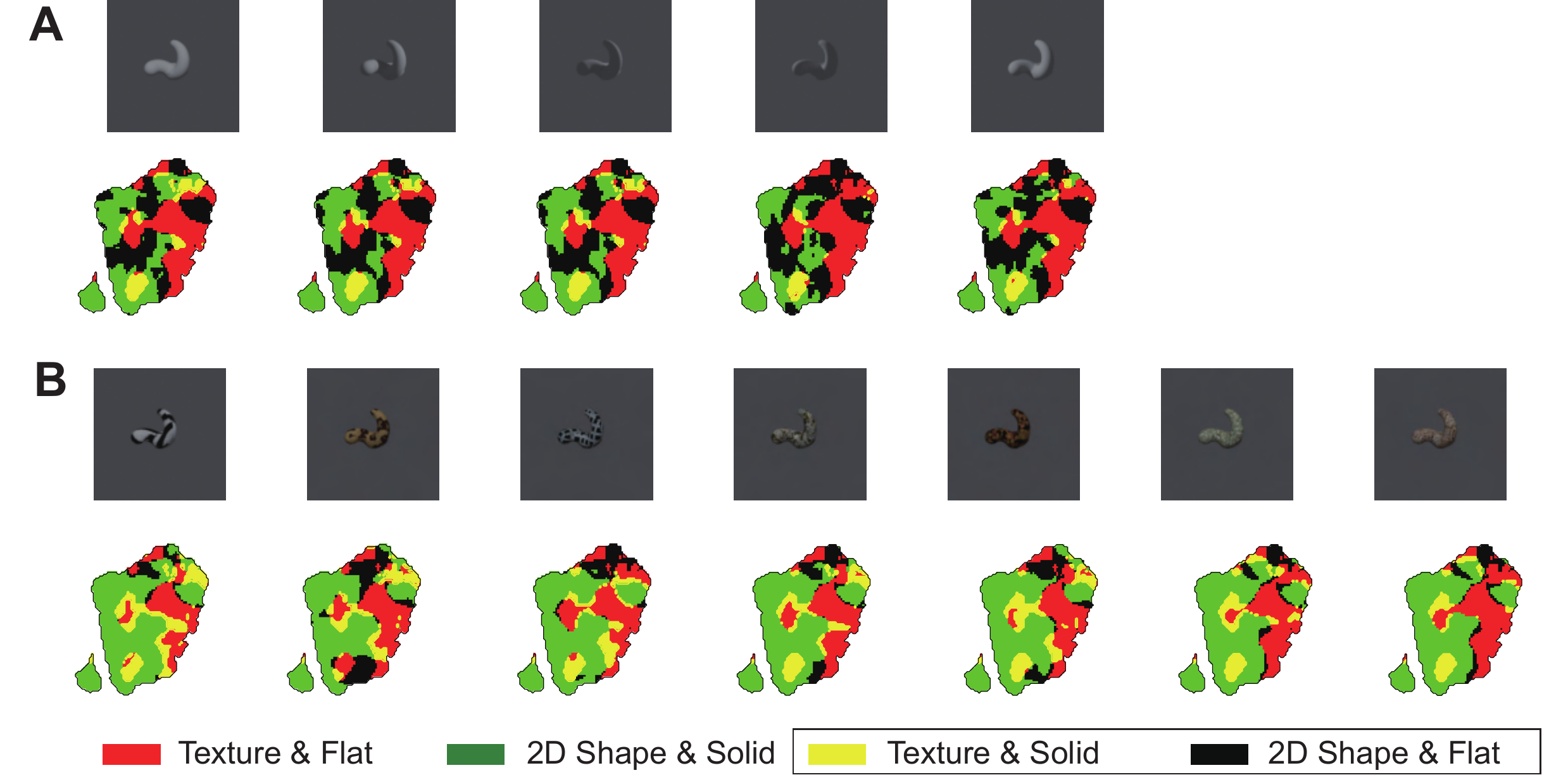}
    \caption{\textbf{Consistency maps across conditions.} Colors mark the four quadrants in Fig.~\ref{fig:scatter}B. Red: texture‐preferring neurons (texture index $>0.5$) with flat preference (SFI $<0$) — \emph{consistent}. Green: 2D‐shape–preferring neurons (texture index $<0.5$) with solid preference (SFI $>0$) — \emph{consistent}. Yellow: texture‐preferring with solid preference (SFI $>0$) — \emph{inconsistent}. Black: 2D‐shape–preferring with flat preference (SFI $<0$) — \emph{inconsistent}. \textbf{A}. Consistency across shading directions. \textbf{B}. Consistency across texture families.}
    \label{fig:appendix_consistent_map}
\end{figure*}

\begin{figure*}[!t]
    \centering
    \includegraphics[width=\linewidth]{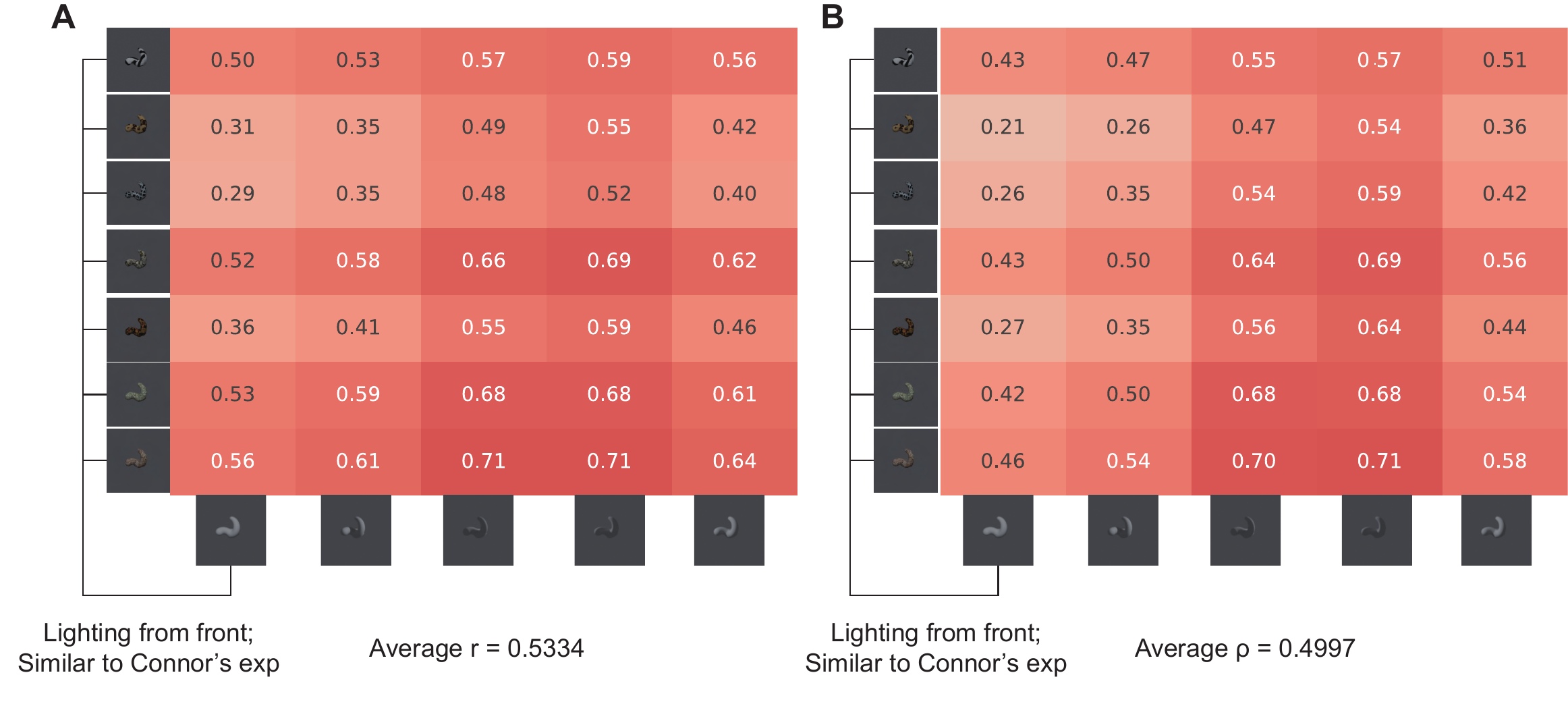}
    \caption{Heatmaps of correlations between SFI measured under shading and texture renderings. Each column corresponds to a shading condition and each row to a texture condition. \textbf{(A)} Pearson correlation coefficients. \textbf{(B)} Spearman rank correlations.}
\label{fig:appendix_shade_texture_SFI_corr}
\end{figure*}

\begin{figure*}[!t]
    \centering
    \includegraphics[width=\linewidth]{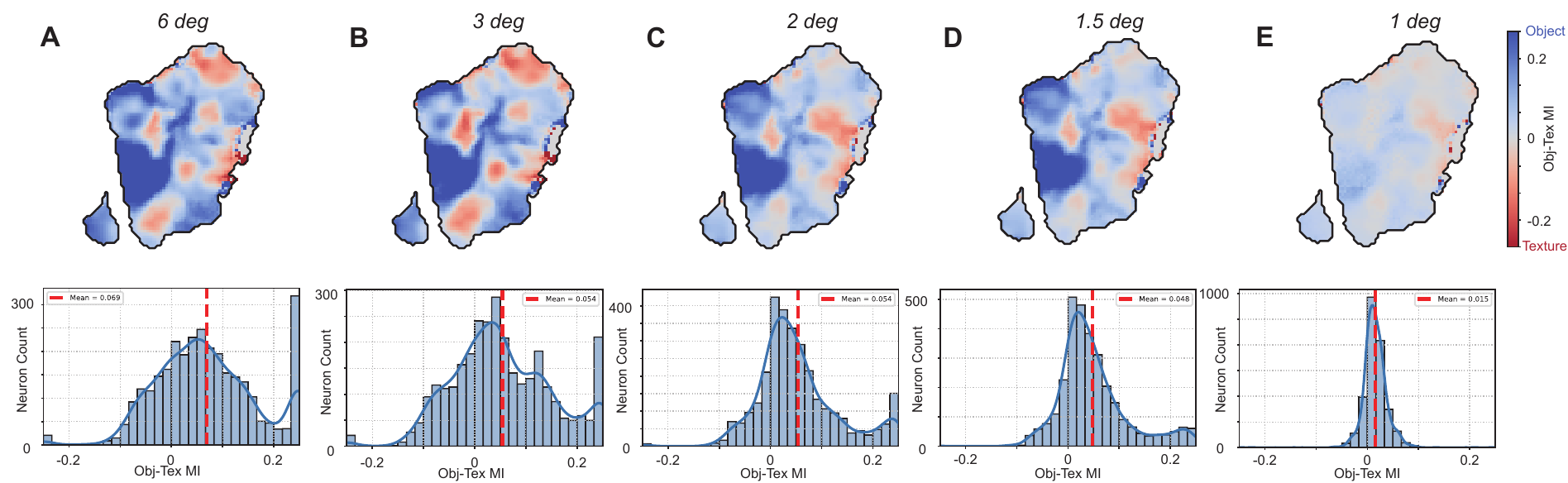}
    \caption{
    \textbf{Topographic organization of the Object--Texture Modulation Index (MI) across $6^{\circ}$--$1^{\circ}$
.}
    \textbf{(A)} Object-Texture Modulation based on $6^{\circ}$: MI map and corresponding distribution of MI values.
    \textbf{(B)} Object-Texture Modulation based on $3^{\circ}$: MI map and corresponding distribution of MI values.
    \textbf{(C)} Object-Texture Modulation based on $2^{\circ}$: MI map and corresponding distribution of MI values.
    \textbf{(D)} Object-Texture Modulation based on $1.5^{\circ}$: MI map and corresponding distribution of MI values. 
    \textbf{(E)} Object-Texture Modulation based on $1^{\circ}$: MI map and corresponding distribution of MI values.
    MI ranges from $-1$ to $1$; negative values indicate a preference for Portilla \& Simoncelli scrambled texture stimuli \cite{Portilla2000}, whereas positive values indicate a preference for the object.
    Second row: The distribution of the MI values in the four cases. }
    \label{fig:appendix_tex_obj}
\end{figure*}

\end{document}